\documentclass[%
 reprint,
superscriptaddress,
 amsmath,amssymb,
 aps,
prb,
]{revtex4-1}

\usepackage{mathtools}
\usepackage{graphicx}% Include figure files
\usepackage{subcaption}% get subfigures
\usepackage{dcolumn}% Align table columns on decimal point
\usepackage{bm}% bold math
\usepackage{color}

\begin{document}

%\preprint{APS/123-QED}

\title{Dipolar Spin Ice Under Uniaxial Pressure}% Force line breaks with \\

\author{R. Edberg}
 \affiliation{Physics Department, KTH Royal Institute of Technology, Sweden}%Lines break automatically or can be forced with \\
 \author{L. \O rduk Sandberg}
 \affiliation{Nanoscience Center, Niels Bohr Institute, University of Copenhagen, Universitetsparken 5, DK-2100 Copenhagen \O , Denmark}%Lines break automatically or can be forced with \\
 
\author{I. M. Bergh Bakke}
\affiliation{University of Oslo, Centre for Materials Science and Nanotechnology, NO-0315, Oslo, Norway}%Lines break automatically or can be forced with \\

\author{M. L. Haubro}
\affiliation{Nanoscience Center, Niels Bohr Institute, University of Copenhagen, Universitetsparken 5, DK-2100 Copenhagen \O , Denmark}%Lines break automatically or can be forced with \\ 

\author{L. C. Folkers}
\affiliation{Department of Chemistry, University of Lund, Sweden}

\author{L. Mangin-Thro}
\author{A. Wildes}
\affiliation{Institut Laue-Langevin, 38042 Grenoble, France}

\author{O. Zaharko}
\affiliation{Laboratory for Neutron Scattering, Paul Scherrer Institute, 5232 Villigen-PSI, Switzerland}

\author{M. Guthrie}%
 \affiliation{European Spallation Source ERIC, 22363 Lund, Sweden}
%forced with \\
\affiliation{University of Edinburgh, School of Physics and Astronomy and Centre for Science at Extreme Conditions}%Lines break 

\author{A. T. Holmes}%
 \affiliation{European Spallation Source ERIC, 22363 Lund, Sweden}
 
\author{M. H. S{\o}rby}%
\affiliation{Department for Neutron Materials Characterization, Institute for Energy Technology, P.O. Box 40, NO-2027, Kjeller, Norway}%Lines break automatically or can be forced with \\

%\author{H. Fjellv\r ag}%
%\affiliation{Chemistry Department, University of Oslo, Norway}%Lines break automatically or can be forced with \\

\author{K. Lefmann}%
\affiliation{Nanoscience Center, Niels Bohr Institute, University of Copenhagen, Universitetsparken 5, DK-2100 Copenhagen \O , Denmark}%Lines break automatically or can be forced with \\

\author{P. P. Deen}%
\affiliation{Nanoscience Center, Niels Bohr Institute, University of Copenhagen, Universitetsparken 5, DK-2100 Copenhagen \O , Denmark}%Lines break automatically or can be 
\affiliation{European Spallation Source ERIC, 22363 Lund, Sweden}
%forced with \\

\author{P. Henelius}%
\affiliation{Physics Department, KTH Royal Institute of Technology, Sweden}
\affiliation{Faculty of Science and Engineering,  \r{A}bo Akademi University, \r{A}bo, Finland}%Lines break automatically or can be forced with \\

 %\email{ricedb@kth.se}

\date{\today}% It is always \today, today,

\begin{abstract}
The magnetically frustrated spin ice family of materials is host to numerous exotic phenomena such as magnetic monopole excitations and macroscopic residual entropy extending to low temperature. A finite-temperature  ordering transition in the absence of applied fields has not been experimentally observed in the classical spin ice materials Dy$_2$Ti$_2$O$_7$ and  Ho$_2$Ti$_2$O$_7$. Such a transition could be induced by the application of pressure, and in this work we consider the effects of uniaxial pressure on classical spin ice systems. Theoretically we find that the pressure induced ordering transition in Dy$_2$Ti$_2$O$_7$ is strongly affected by the dipolar interaction. We also report measurements on the neutron structure factor of Ho$_2$Ti$_2$O$_7$ under pressure, and compare the experimental results to the predictions of our theoretical model. \end{abstract}

\pacs{Valid PACS appear here}% PACS, the Physics and Astronomy
\maketitle

%\tableofcontents

\section{Introduction}

In a frustrated magnet a unique ground state can be difficult to reach, since the primary interactions are obstructed by competition between interactions of similar strength or a special geometry of the lattice, leading to a strong ground state degeneracy in the unperturbed case\cite{Springer_frust_book}.   Therefore, an eventual low-temperature ordering transition and resolution of the third law of thermodynamics is often driven by weak, perturbative interactions whose effects are typically hidden in non-frustrated systems. In recent years, the geometrically frustrated spin ice compounds have revealed many unusual phenomena such as magnetic monopole excitations \cite{Castelnovo_Nature}, fractionalization of the magnetic moment \cite{Brooks} and a low-temperature Coulomb phase \cite{Henley_Coulomb}. However, a low-temperature  ordering transition below the Coulomb phase in the absence of applied fields  has not been experimentally observed in the classical spin ice compounds Dy$_2$Ti$_2$O$_7$ (DTO) and  Ho$_2$Ti$_2$O$_7$ (HTO) \cite{Poma13,henelius16}. Theoretically there are predictions for a dipolar-driven low-temperature transition in the $0.2 \textup{ K}$ temperature range \cite{MelkoGingras}, but experimentally the measurements are challenging due to extremely slow relaxation effects \cite{Poma13}, and the nature of such a transition is not known.

One way to increase the possible transition temperature into a more experimentally accessible range is to apply a suitable perturbation, such as pressure, to the system. Previous experimental studies detect a clear uniaxial pressure-induced signature in the susceptibility of DTO \cite{MITO}, while hydrostatic pressure has no noticeable effect on HTO \cite{Mirebeau_2004}. A previous theoretical study predicts that uniaxial pressure should give rise to an unusual infinite order phase transition in DTO, with features of both a continuous and discontinuous transition, to a ferromagnetic state \cite{Jaubert1}. In this study, we include the long-range dipolar interaction in the theoretical model, and find that it does not favor the ferromagnetic state. We also perform spin-polarized neutron scattering experiments on HTO under uniaxial pressure and compare the results to our theoretical predictions.  While our model captures the qualitative features of the experiment, further measurements are needed in order to fully understand the effects of pressure on the low-temperature spin ice state in DTO and HTO.

\section{Model, Simulation and Experimental methods}

\subsection{Model}
Existing literature presents a large number of spin ice models, ranging from effective monopole models \cite{Castelnovo_Nature}, electrolyte theories \cite{Bramwell_Nature} to microscopic models incorporating for example dipolar, exchange and even hyperfine interations \cite{hertog00, yavo08, henelius16}. Different models emphasize different aspects of the physical properties of spin ices. We aim to  select the simplest model that captures the most relevant aspects of the pressure dependence of the system.  In the previous theoretical study on the effects of pressure in the classical spin ice materials, a nearest neighbor model was used \cite{Jaubert1}. Due to a self-screening of the long-range part of the dipolar interaction, this relatively simple model gives a surprisingly good description of classical spin ice behaviour \cite{denHertog,Isakov_SS}. It can be analyzed analytically using a Husimi tree solution \cite{jaubert13}, and the low-temperature properties are similar to those of the monopole model \cite{Castelnovo_Nature}. However, to capture the experimental low-temperature behavior of  DTO and HTO, dipolar interactions have proven to be important \cite{hertog00}, and we  therefore consider the dipolar spin ice model. 

The Hamiltonian for normalized spins $\mathbf{S}_i$ residing on the sites $\textbf{r}_i$ of the pyrochlore lattice with nearest neighbor distance $a$, is given by a nearest neighbor antiferromagnetic exchange interaction $J$ and a long-range dipolar interaction $D$, 

\begin{equation}\label{Hamiltonian}
	\begin{split}
		\mathcal{H}=&\sum_{\langle i,j\rangle}J(i,j) \, \mathbf{S}_i\cdot\mathbf{S}_j+\\
		&Da^3\sum_{i<j}\left(\frac{\mathbf{S}_i\cdot\mathbf{S}_j}{|\mathbf{r}_{ij}|^3}-3\frac{\left(\mathbf{S}_i\cdot\mathbf{r}_{ij}\right)\left(\mathbf{S}_j\cdot\mathbf{r}_{ij}\right)}{|\mathbf{r}_{ij}|^5}  \right).
	\end{split}
\end{equation}

\begin{figure}[!h]
	\includegraphics[width=1\linewidth,trim = 0mm 0mm 0mm 0mm, clip]{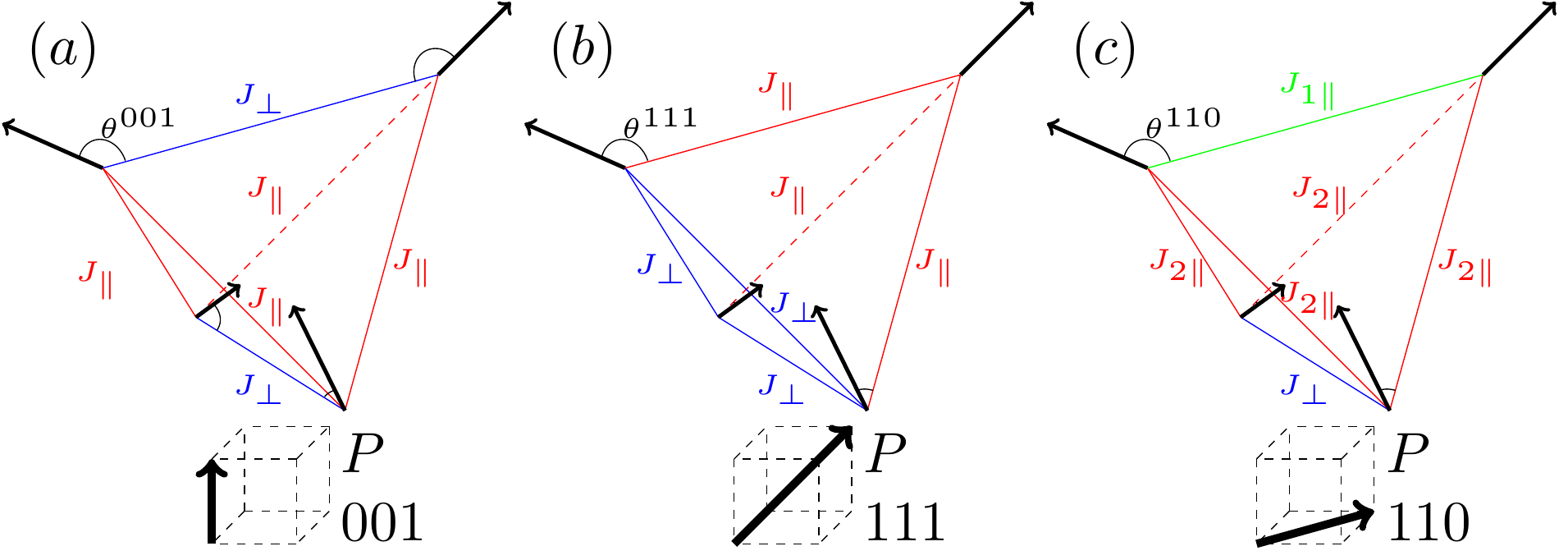}
	\caption{\raggedright Model for uniaxial pressure applied in the  [001](a), [111](b) and [110](c) directions. $J_\perp$ remains fixed while $J_{(k)\parallel}$ vary with pressure. Colors indicate the different exchange interactions.
	The angle $\theta$ between the Ising-axis and the side of the tetrahedron is altered to $\theta^{001}$,$\theta^{111}$ and $\theta^{110}$ when the crystal lattice is compressed in the respective direction.
	}
	\label{model}
\end{figure}

Here, the angle brackets denote summation over nearest neighbors and $\textbf{r}_{ij}=\textbf{r}_i-\textbf{r}_j$. 
The antiferromagnetic interaction $J$ is caused by oxygen-mediated superexchange between adjacent magnetic ions. Due to the strong crystal field, the spins are oriented along the local $[111]$-axis connecting the center points of adjacent tetrahedra. To capture the detailed behavior of a number of physical properties for DTO, exchange interactions up to third nearest neighbors have been determined \cite{yavo08}. For simplicity, we do not investigate the effects of pressure on weaker further neighbor interactions, and truncate the exchange (but not the dipolar) interaction  at the first nearest neighbor. We therefore use as our starting point, the so-called s-DSM model with $J= 3.72 \textup{ K}$, $D=1.41 \textup{ K}$ \cite{MelkoGingras} for DTO and $J= 1.56 \textup{ K}$, $D=1.41 \textup{ K}$ for HTO \cite{Jaubert_2011}.

We model compression of the crystal with zero Poisson ratio, as suggested by experiment \cite{MITO}. When atoms move closer, the wave function overlap and corresponding exchange parameters are expected to change. We make the assumption that $J(i,j)=J_{\perp}=J_{P=0}$ if the neighbors $i$ and $j$ are in a lattice plane perpendicular to the direction of applied pressure and $J(i,j)=J_{k\parallel}$ if they are not. $J_{P=0}$ is the ambient pressure value of the s-DSM model. $k$ labels the $6$ different directions of nearest neighbor bonds and we make no a priori assumptions on the magnitude of $J_{k\parallel}$. The dipolar interactions become stronger along the direction of compression, growing as $|\textbf{r}_{ij}|^{-3}$, when distances decrease. Furthermore, we make the assumption that the spins continue to point towards the centers of the tetrahedra as the crystal deforms, altering the local  $[111]$-axis. This change of spin directions leads to a change in both exchange and dipolar interactions, due to the dot products in Eq.~(\ref{Hamiltonian}). 

Motivated by experiments \cite{MITO}, we model pressure applied along the $[001]$, $[111]$ and $[110]$ directions. By symmetry, for the first two directions, it suffices to model the nearest neighbor interactions with two parameters, $J_\perp$ and $J_\parallel$. For the $[110]$ direction, however, the lower symmetry of the lattice requires us to use three different exchange parameters, $J_\perp$, $J_{1\parallel}$ and $J_{2\parallel}$. We illustrate these different exchange interactions for a single tetrahedron in Fig.~\ref{model} for the three respective directions of pressure. ($J_{(k)\parallel}$ need not be the same for the different directions of pressure.) 

Demagnetizing effects in susceptibility measurements can, under ambient pressure, be minimized by using needle-shaped samples. However, such samples are likely to break under pressure, and less elongated samples are used in practice. It is therefore important to consider the demagnetizing transformation
\begin{equation}\label{eq:demagtrans}
B_\textup{ext}=B_\textup{int}+N M.	
\end{equation} 
which gives the relationship between the internal field, $B_\textup{int}$, the external field, $B_\textup{ext}$, the magnetization, $M$, and the demagnetizing factor of the sample, $N$. Note that $N$ depends on the shape of the sample and therefore changes upon compression. We find that this effect cannot be ignored in an accurate description of the experiments \cite{MITO}. Hence the uniaxial pressure directly alters both microscopic properties; such as spin directions and interaction parameters, and macroscopic properties; such as the magnetization and shape of the sample.

In previous magnetization measurements, it was found that the signatures of pressure is largest when pressure is applied along the $[001]$ direction \cite{MITO}. Earlier theoretical work predicts a low-temperature transition into an ordered phase carrying net magnetization along this axis \cite{Jaubert1}. This motivates us to make an extended study of the zero temperature phases for our model in the $[001]$ direction to investigate the system at low temperature.

\subsection{Simulation method and demagnetization corrections}

Monte Carlo simulations using single spin flip and loop updates are used to investigate a number of different system sizes with periodic boundary conditions according to the Metropolis-Hastings algorithm. We use the 16 particle standard cubic unit cell \cite{MelkoGingras}. All super-cells used are cubic of size $L^3$, $L\in[1,\cdots,8]$. Ewald summation is used to effectively account for the long range conditionally convergent dipolar contributions \cite{frenkel2001understanding}. The geometrical compression is introduced by an appropriate linear coordinate transformation depending on the direction of applied pressure. In particular, the Ewald summation is done using sheared lattice vectors for the reciprocal space contribution. 

In order to account for the macroscopic boundary effects, demagnetization transformations are made by modeling the sample as a prolate ellipsoid. The major axes of the ellipsoid are given by the dimensions of the samples used in the experiment. For the DTO experiment\cite{MITO}, only the shape of the $[001]$ and $[110]$ crystals could be recovered; $1.77\times 1.80 \times 1.61 \textup{ mm}^3$ ($[001]$ pressure along the side of length $1.61\textup{ mm}$), and  $1.28\times 0.96 \times 1.63\textup{ mm}^3$ ($[110]$ pressure along the side of length $1.63 \textup{ mm}$) \cite{Priv_Comm_Mito}. These dimensions give $N_{001}=0.319$ and $N_{110}=0.426$ respectively in the high-temperature limit \cite{Osborn1945}, and we have not taken the temperature dependence of the demagnetization factor into account \cite{Micke17}. The sample dimensions are deformed in accordance with the microscopic compression of the lattice along the direction of pressure. The dimensions for the sample measured in the $[111]$ direction were unavailable and we assume the shape to be a perfect sphere with $N_{111}=1/3$ under ambient pressure. 

\subsection{Experimental Methods}

Magnetic neutron diffraction experiments were performed at the Institut Laue Langevin (ILL) using the polarized diffuse scattering instrument, D7 \cite{D7, dataThatWeTookOnILL}. Elastic neutron diffraction profiles with uniaxial polarization analysis (neutron polarization along the vertical axis) were measured for HTO under uniaxial pressure with an incident wavelength $\lambda$ = 4.86 \AA{}.  Polarisation analysis on D7 measures spin flip and non-spin flip contributions. In the limit of negligible background and nuclear spin incoherent scattering, the spin flip scattering will have only the magnetic scattering given by Eq.~(\ref{Eq.SFstructureFactor}). The data have been corrected for detector and polarization analyzer efficiencies using standard samples of vanadium and amorphous silica respectively \cite{D7}.

Uniaxial pressure was applied to HTO crystals using a Cu-Be anvil-type pressure cell optimized for uniaxial pressure (details of the cell will be published elsewhere \cite{cell}).  Temperature control was achieved using a liquid helium cryostat. The samples were cylindrical shaped single crystals, $V = 10$~mm$^{3}$, cut from the larger image-furnace-grown crystals, with a surface alignment better than $0.1^{\circ}$. The pressure applied was determined via force gauge calibrations considering low temperature changes to the cell with a maximum pressure of 2.2(0.5)~GPa.

The crystals were synthezised using an image furnace located at the Univerity of Lund with ambient oxygen conditions. Structural neutron diffraction measurements on ZEBRA of the Paul Scherrer Institute, Switzerland, showed an absence of structural diffuse scattering, which would have indicated oxygen defect clusters \cite{Sala14}.

\section{Theoretical and Experimental Results}

\subsection{Magnetization}
\begin{figure}[!h]
	\centering
	\includegraphics[width=1\linewidth]{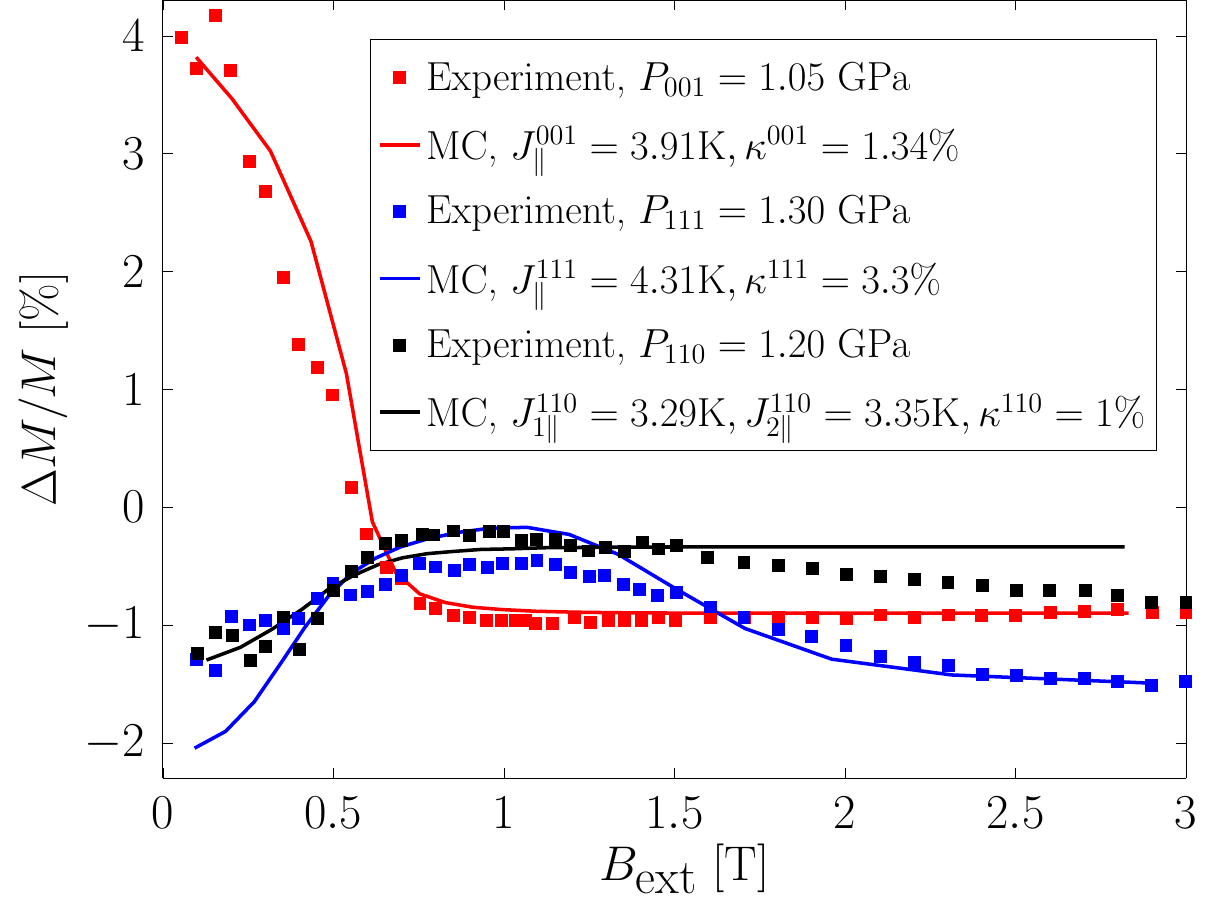}
	\caption{\raggedright Relative increase in the sample magnetization $\Delta M/M=(M_{P>0}-M_{P=0})/M_{P=0}$ when field and pressure are both applied in the $[001]$, $[111]$ and $[110]$ directions at $1.7\textup{ K}$. Experimental result\cite{MITO} (dotted) and our fit by MC simulation for a system of 8192 spins (solid).}
	\label{fig:dM/Mcurves}
\end{figure}

The magnetization of DTO single crystals under applied uniaxial pressure and magnetic field has earlier been measured at $T= 1.7\textup{ K}$ \cite{MITO}. The magnetic field ranged up to $3 \textup{ T}$ and was parallel to the axis of pressure. Measurements were performed with pressure along the $[001]$, $[111]$ and $[110]$ directions. We adjust the parameters of the proposed model, Eq.~(\ref{Hamiltonian}), to match the experimental data. The best match between experiment and model is shown in Fig.~\ref{fig:dM/Mcurves}, and next we describe the details of the modeling procedure. 

We assume that the magnetic moments retain their Ising symmetry under pressure, and that a sufficiently strong field aligns the moments. For such a field, changes in the demagnetization factor or in the interaction energies will have no effect on the magnetization. The difference in saturation magnetization observed when applying pressure in different directions is purely due to lattice compression which changes the angle of the local [111]-axis. The components of the magnetic moments parallel to the direction of the applied field decreases, which reduces the saturation magnetization along the relevant direction. From the value of the saturation magnetization we can therefore determine the compression $\kappa$. We find that the crystals deform by $\kappa^{001}=1.34\%$, $\kappa^{111}=3.3\%$ and $\kappa^{110}=1.0\%$ for the $[001]$, $[111]$ and $[110]$ directions respectively. This gives the demagnetization factors $N_{001}^\kappa=0.322$, $N_{110}^\kappa=0.433$ and $N_{111}^\kappa=0.341$ under compression, for the respective directions.

The data for the $[001]$ direction appears saturated already at about $1$ T, while we assume that the data for the $[111]$ direction has saturated at 3 T. For the $[110]$ direction the experimental data has not saturated at high field. We find this result peculiar since the configuration that minimizes the Zeeman energy can be reached without leaving the spin ice manifold for fields along $[110]$ and $[001]$, while this is not so for the $[111]$ direction. We would expect that a stronger field is needed  to create the necessary monopole excitations for full $[111]$ saturation, but this is not reflected by the experiment. Due to this anomaly in the high field behavior we take the value at $1.5 \textup{ T}$, which would firmly saturate the crystal in the $[001]$ direction, to be the saturation value for the $[110]$ direction, resulting in a compression of $1\%$.

\begin{figure}[!h]
	\centering
	\includegraphics[width=1\linewidth]{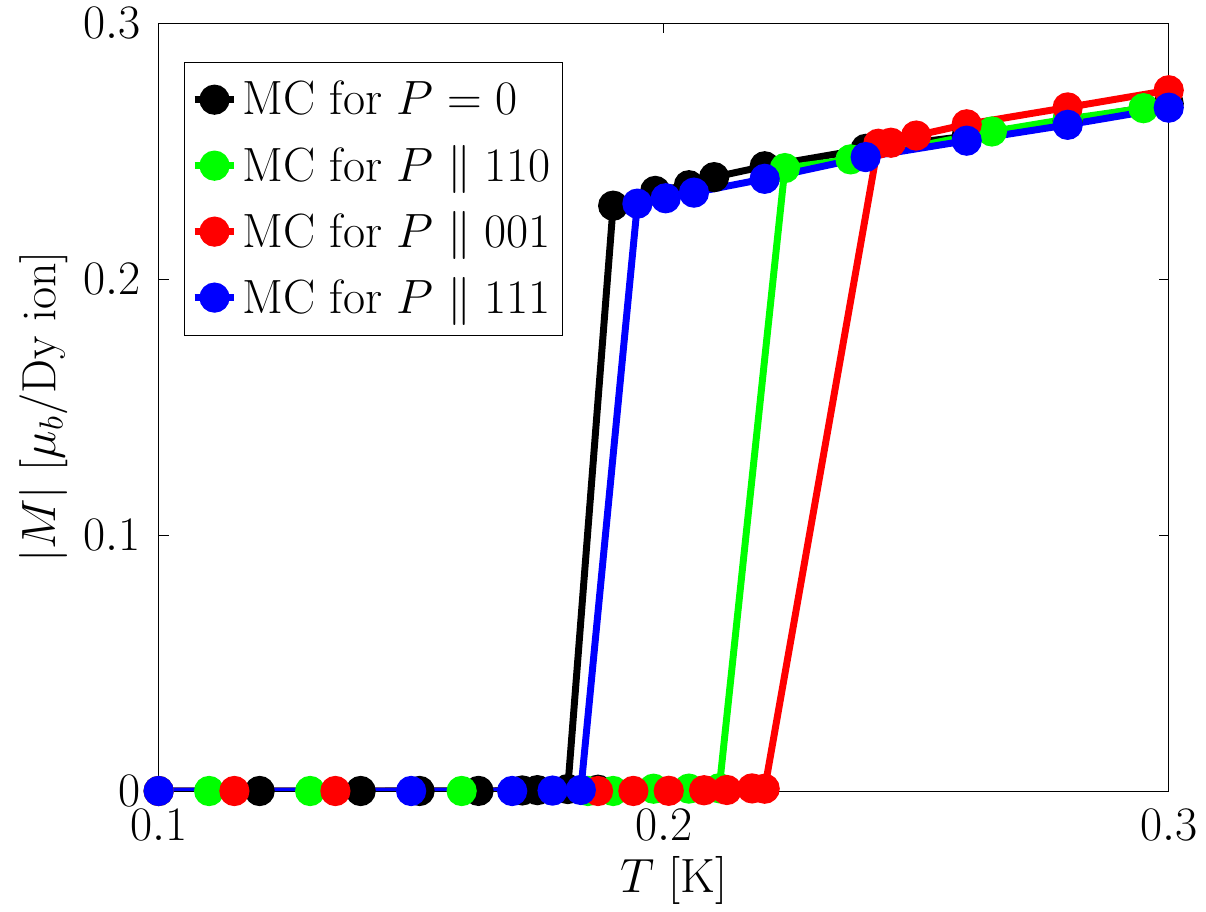}
	\caption{\raggedright Simulated values of magnetization plotted against temperature for 1024 spins. Ambient pressure (black) together with uniaxial pressure along $[110]$ ($1.20$ GPa, green), $[001]$ ($1.05$ GPa, red) and $[111]$ ($1.30$ GPa, blue).}
	\label{fig:spinChainState}
\end{figure}

With the compression and demagnetization factor fixed, the $J_\parallel$ exchange parameter is adjusted in order to give the best fit for the measured magnetization as a function of external field; these data are shown as continuous lines in Fig.~\ref{fig:dM/Mcurves}. It was found that for both the $[001]$ and $[111]$ directions, the $J_\parallel$ coupling increases from the original value of $J_\parallel=3.72$ K. Our best estimate gives $J^{001}_\parallel=3.91 \textup{ K}$ and $J^{111}_\parallel=4.31 \textup{ K}$. For the $[110]$ direction, the parameters decrease and it is the difference $J^{110}_{1\parallel}-J^{110}_{2\parallel}$ that determines the zero field limit in magnetization change. Our best fit gives $J^{110}_{1\parallel}= 3.29 \textup{ K}$ and $J^{110}_{2\parallel}=3.35 \textup{ K}$. Assuming that the exchange parameters are independent of temperature we find that, for all directions, the ground state of our model is a dipolar spin chain state with zero net magnetization, which is also the expected ground state of the dipolar spin ice model under ambient pressure\cite{MelkoGingras, henelius16}.

\begin{figure}[!h]
	\centering
	\includegraphics[width=1\linewidth]{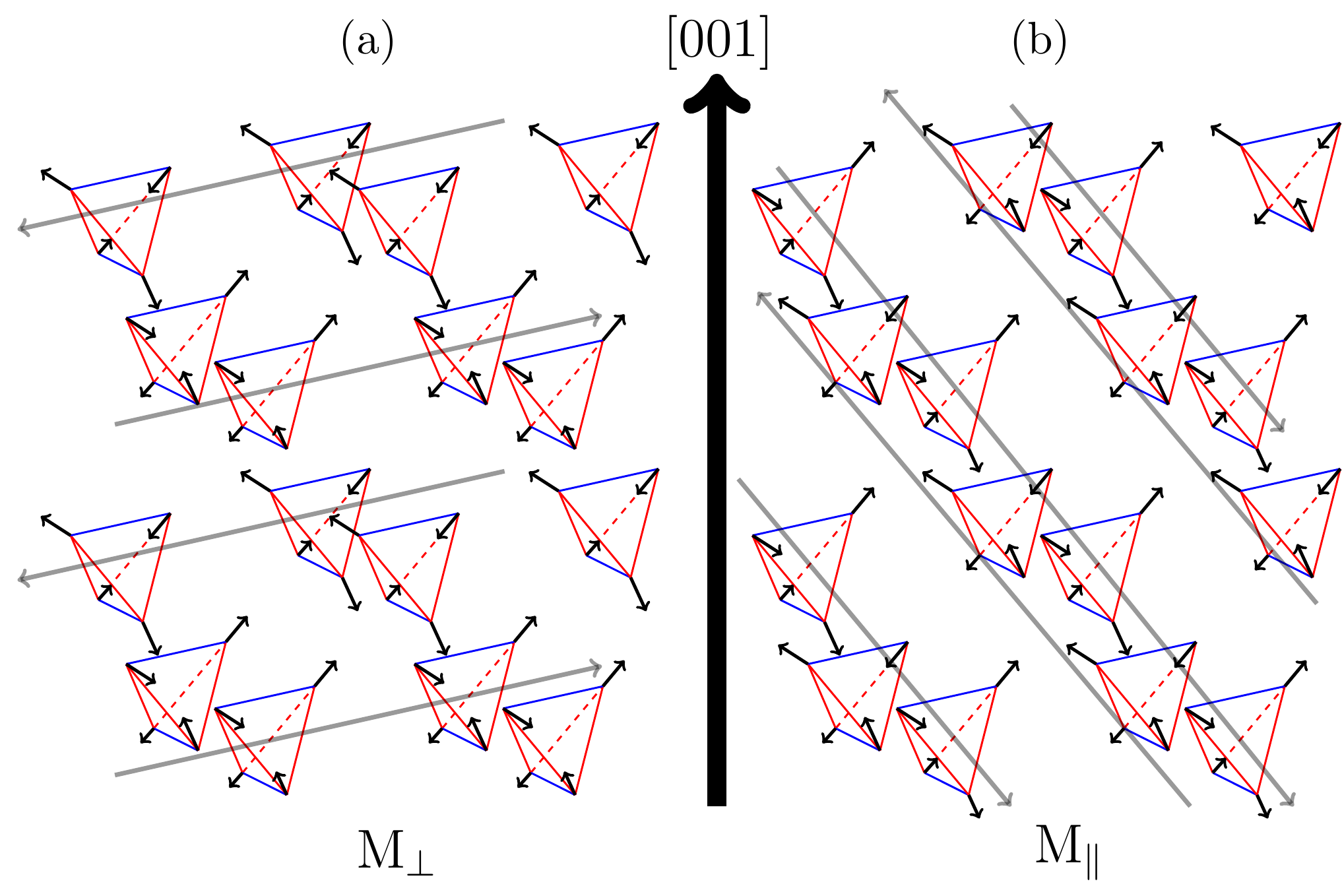}
	\caption{\raggedright The degeneracy is lifted under application of pressure. For the $[001]$ direction, there exists two submanifolds of ground states; (a) $M_\perp$ and (b) $M_\parallel$. Red and blue bonds corresponds to $J_\parallel$ and $J_\perp$ respectively. Only up-tetrahedrons are drawn for clarity. Grey lines indicate the directions of the spin chains.}\label{fig:MMStates}
\end{figure}

This is illustrated in Fig.~\ref{fig:spinChainState}, which shows how the finite size absolute magnetization drops to zero in a first order transition. The transition temperature increases with pressure,  especially when pressure is applied in the $[001]$ and $[110]$ directions. The twelve-fold degeneracy of the ground state is lifted due to the change of symmetry in the deformed unit cell. With pressure along $[001]$, the ground state manifold is split into two submanifolds. One with chains perpendicular to the $[001]$ direction, $\textup{M}_\perp$,  and one with chains having a component along $[001]$, $\textup{M}_\parallel$, see Fig.~\ref{fig:MMStates}. Our fit to magnetic susceptibility data for DTO, finds that the $\textup{M}_\parallel$ state has the lowest energy of the two possible states.

\subsection{Ground states for pressure along $[001]$}
\begin{figure}[!h]
	\centering
	\includegraphics[width=1\linewidth]{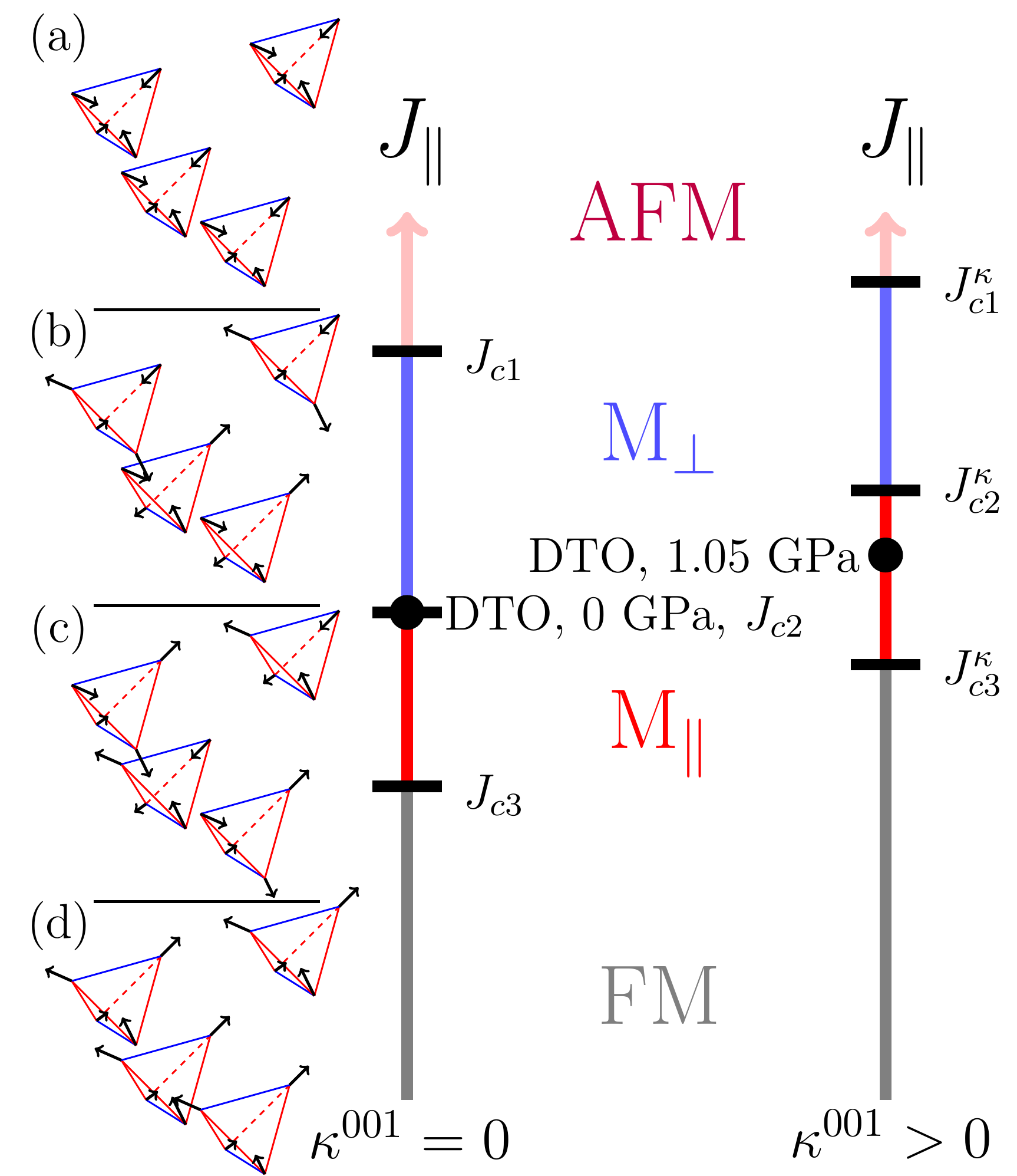}
	\caption{\raggedright Ground states of the model schematically plotted with varying $J_\parallel$ in the $[001]$ direction. The values $J_{c1},J_{c2},J_{c3}$ and $J_{c1}^\kappa,J_{c2}^\kappa,J_{c3}^\kappa$ denote the phase boundaries without and with compression respectively. We plot instances of the ground states in the leftmost column for the 16 particle unit cell. Red and blue bonds corresponds to $J_\parallel$ and $J_\perp$ respectively. Only up-tetrahedrons are drawn for clarity.}\label{fig:different[001]phases}
\end{figure}

 In Fig.~\ref{fig:different[001]phases}, we illustrate schematically the ground states for our spin ice model, Eq.~(\ref{model}), when varying $J_\parallel$ and lattice compression for pressure applied in the $[001]$ direction. We find that there are four possible ground states: (a) the all-in-all-out state (AFM), (b),(c) The spin chain states ($\textup{M}_\perp$, $\textup{M}_\parallel$) and (d) the earlier proposed ferromagnetic state (FM). We introduce the critical values $J_{c1},J_{c2},J_{c3}$ and $J_{c1}^\kappa,J_{c2}^\kappa,J_{c3}^\kappa$ denoting the phase boundaries without and with compression respectively. For DTO without compression, the state boundaries are at $J_{c1}=9.02\textup{ K}$, $J_{c2}=3.72\textup{ K}$, $J_{c3}=3.33\textup{ K}$. For DTO at $1.05 \textup{ GPa}$, inducing $\kappa^{001} = 1.34\%$ compression, we have $J_{c1}^\kappa=9.21 \textup{ K}$, $J_{c2}^\kappa=4.02 \textup{ K}$, $J_{c3}^\kappa=3.64 \textup{ K}$.

In previous work \cite{Jaubert1}, spin ice was modeled for pressure in the $[001]$ direction with effective nearest-neighbor interactions, $\mathcal{H}=-\sum_{\langle i,j\rangle} J_{\textup{eff}}(i,j) \, \mathbf{S}_i\cdot\mathbf{S}_j$, but without dipolar interactions and no lattice compression in the Hamiltonian. It was found by fitting to the experimental data that $J_{\parallel\textup{eff}} > J_{\perp \textup{eff}}$, and that this results in a ferromagnetic phase transition at low temperatures.

\begin{figure}
	\centering
	\begin{subfigure}[b]{0.23\textwidth}
	    \includegraphics[width=\textwidth]{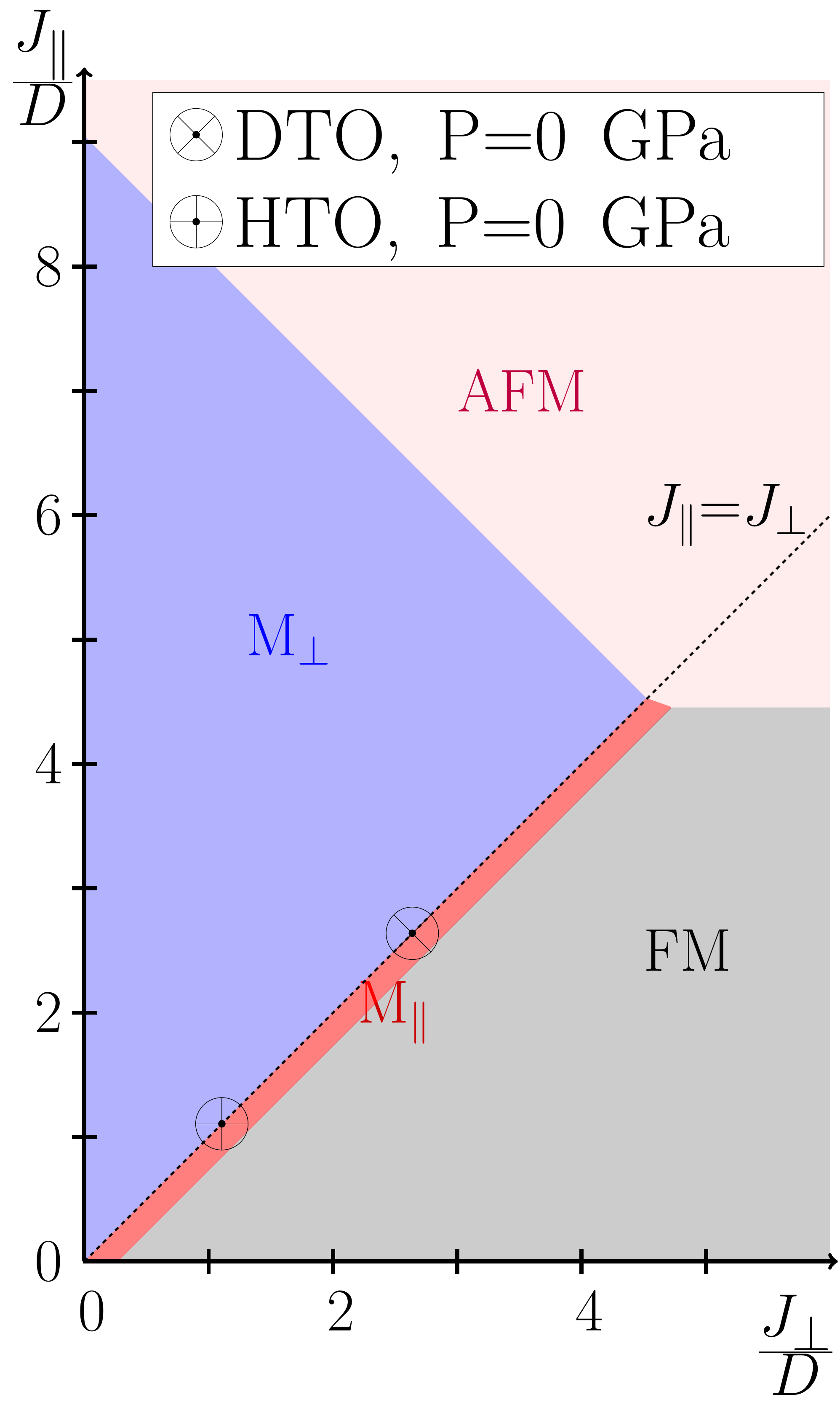}
	    \caption{$\kappa^{001}=0$}\label{fig:NoDefPhaseDiagram}
	\end{subfigure}
	\begin{subfigure}[b]{0.23\textwidth}
	    \includegraphics[width=\textwidth]{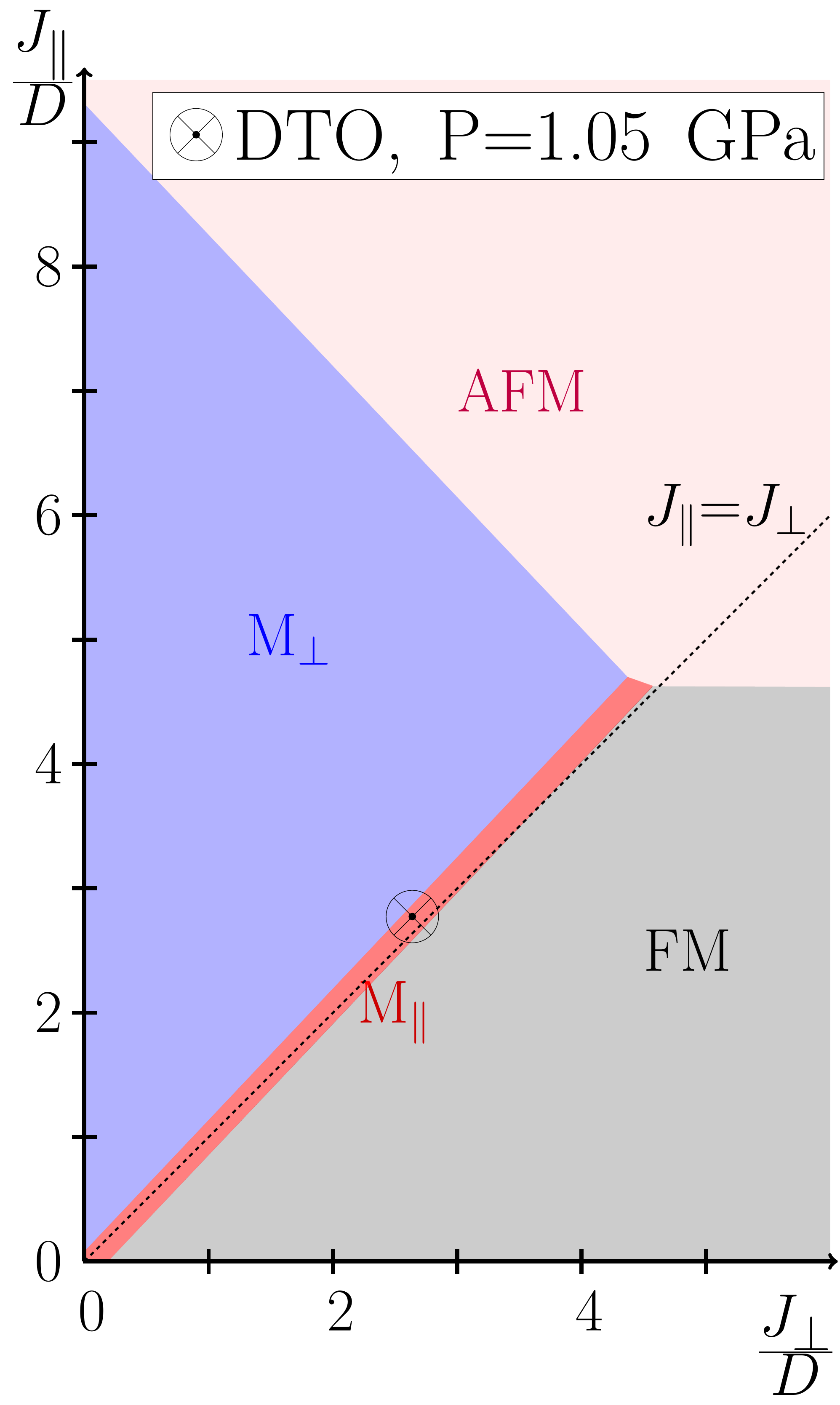}
	    \caption{$\kappa^{001}=1.34\%$}\label{fig:defPhaseDiagram}
	\end{subfigure}
	\caption{\raggedright Ground state phase diagram in terms of normalized interactions, $J/D$. Results are shown for both zero and $1.34\%$ compression; the value suggested by the DTO magnetization data. Blue, light red, and grey regions represent the $M_\perp$, AFM, and FM states, respectively. The $M_\parallel$ region is very narrow and shown in dark red close to the $J_\perp = J_\parallel$ line. The phase boundaries shift with compression and $J_\parallel$ increases with pressure. We include also the ambient pressure value for HTO.}\label{fig:phaseDiagram}
\end{figure}

In this work, we find that with the inclusion of dipolar interactions the picture is more complicated, since the $M_\parallel$ state will compete with the FM state. The outcome of this is that a larger difference between $J_{\parallel\textup{eff}}$ and $J_{\perp\textup{eff}}$ is needed for the ferromagnetic transition to occur. At the nearest neighbor distance the interaction in the s-DSM model is $J_\textup{eff}= 5D-J$, and we note that in order for $J_\textup{eff}$ to increase, $J$ must decrease. We find that the ground state of the system is FM  only when $J_{\parallel\textup{eff}}>5D - J_{c3} > 5D- J_\perp= J_{\perp\textup{eff}}$, instead of immediately when $J_{\parallel\textup{eff}}>  J_{\perp\textup{eff}}$. Furthermore, with the inclusion of lattice compression and the associated change of the local $[111]$-axis, we find, when fitting to the magnetization measurements, that $J_{\parallel}$ increases as a function of pressure, rather than decreases. This change seemingly moves away from, rather than towards, the FM boundary, as illustrated in the left hand side of Fig.~\ref{fig:different[001]phases}. However, a further complication is that the critical value for the phase boundary $J^\kappa_{c3}$ also increases with lattice compression (right hand side of Fig.~\ref{fig:different[001]phases}). So it may still be possible for the system to display the ferromagnetic transition, but this would require much higher pressure than what has been applied in the current measurements. By linear extrapolation, $J_\parallel$ for DTO would cross the zone boundary at $3.4\textup{ GPa}$ and the FM transition would occur at low temperatures.

In Fig.~\ref{fig:phaseDiagram} we show the general ground state phase diagram as a function of normalized $J_\perp$ and $J_\parallel$ interactions.  Fig.~\ref{fig:NoDefPhaseDiagram} depicts the ground states of the system with no lattice compression. In this case $J_\parallel=J_\perp$ defines the boundary between the $\textup{M}_\perp$ and $\textup{M}_\parallel$ phases. 
When adding compression, the phase boundaries are shifted, which is illustrated as we plot the phase diagram for the experimentally relevant compression in Fig.~\ref{fig:defPhaseDiagram}. In particular, the $\textup{M}_\parallel$ region moves, breaking the $M_\perp-M_\parallel$ degeneracy for $J_\parallel=J_\perp$ in favor of the $\textup{M}_\parallel$ states. The overall behavior is the same, with an AFM ground state for large $J_\parallel$ and a FM ground state for small $J_\parallel$.

\subsection{Prediction of neutron diffraction results for DTO}

Neutron scattering is an ideal and unique tool to extract the spin-spin correlations in materials  \cite{Marshall71}.
This enables a close comparison of experimental measurements  and theoretically calculated correlations, a combination which often plays a crucial role in the understanding of magnetic materials at a microscopic level.

Using parameter values relevant for DTO, we sample the thermal average of the simulated spin-spin correlations and calculate the magnetic structure factor for neutron diffraction, $S_{\textup{mag}}$, according to
\begin{equation}\label{Eq.structureFactor}
S_{\textup{mag}}(\textbf{Q})= \frac{[f(|\textbf{Q}|)]^2}{N}\sum_{ij}\left\langle \textbf{S}^{\perp}_i\cdot \textbf{S}^{\perp}_j\right\rangle e^{i\textbf{Q}\cdot \textbf{r}_{ij}},
 \end{equation}
 where $\textbf{Q}$ is the scattering wave-vector, $N$ is the number of spins in the simulation cell, $f(|\textbf{Q}|)$ is the magnetic form factor, and the spin perpendicular component is given by $\textbf{S}_i^\perp=\textbf{S}_i-\textbf{S}_i\cdot\textbf{Q}/|\textbf{Q}|$. We calculate also the structure factor for magnetic spin flip scattering, relevant for experiments with neutron polarization analysis $S_{\textup{SF}}$ according to
\begin{equation}\label{Eq.SFstructureFactor}
%\begin{gather*}
\begin{split}
&S_{\textup{SF}}(\textbf{Q})=\frac{[f(|\textbf{Q}|)]^2}{N}\\
&\times\sum_{ij}\left\langle \textbf{S}^{\perp}_i\cdot \textbf{S}^{\perp}_j-(\textbf{S}_i\cdot \textbf{P})(\textbf{S}_j \cdot \textbf{P})\right\rangle e^{i\textbf{Q}\cdot \textbf{r}_{ij}},
\end{split}
%\end{gather*}
\end{equation}
where $\textbf{P}$ is the normalized polarization direction of the incident neutron beam, and $\textbf{P}\perp\textbf{Q}$ \cite{Blume}. In our experiment, the neutron polarization direction is parallel to the direction of applied pressure and we calculate the neutron scattering profiles for the plane in reciprocal space perpendicular to the direction of the applied pressure.

\begin{figure}[!h]
	\centering
	\begin{subfigure}[b]{0.22\textwidth}
	    \includegraphics[width=\textwidth]{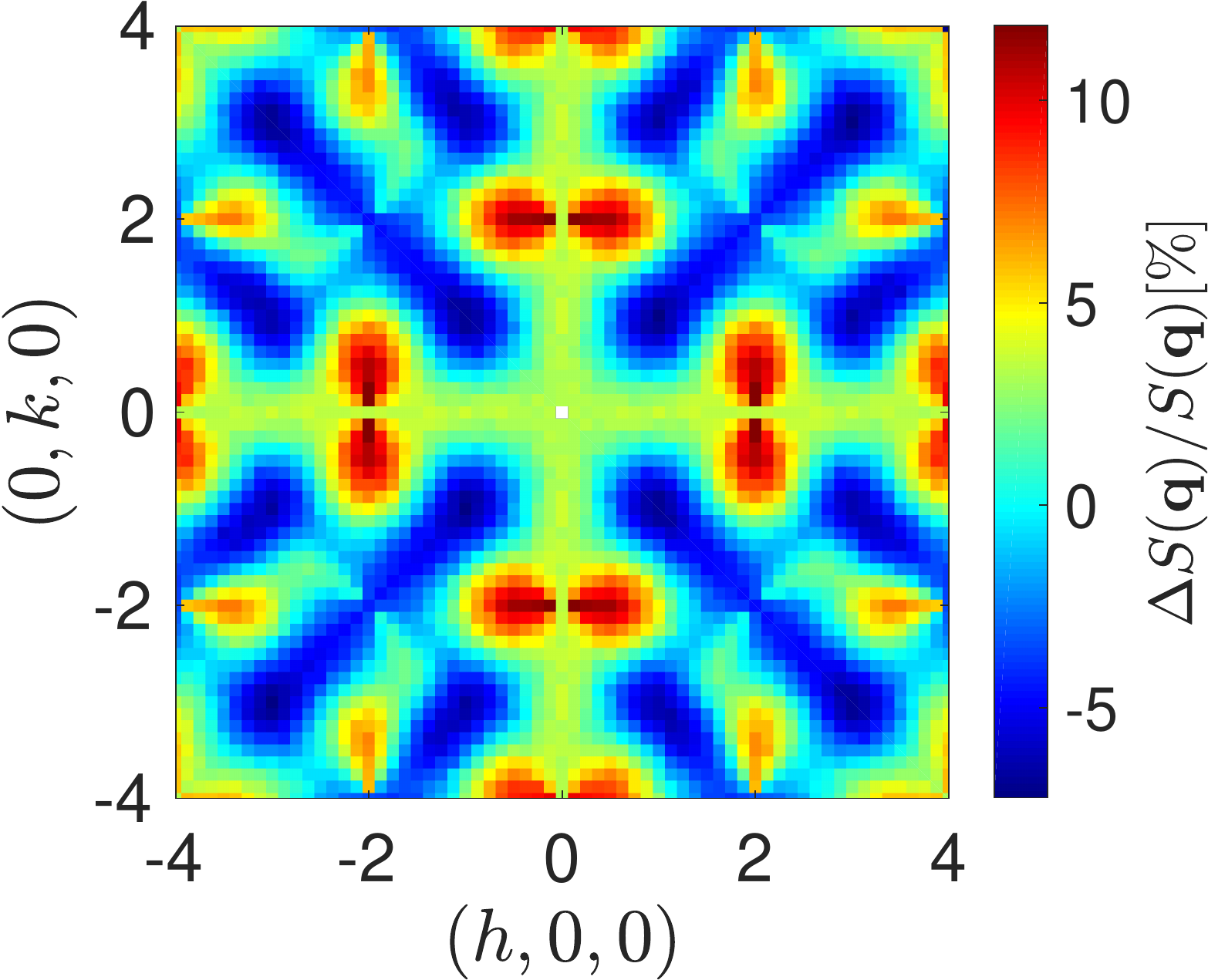}
	    \caption{$[001]$ Magnetic.}\label{[001]Total}
	\end{subfigure}
	\begin{subfigure}[b]{0.22\textwidth}
	    \includegraphics[width=\textwidth]{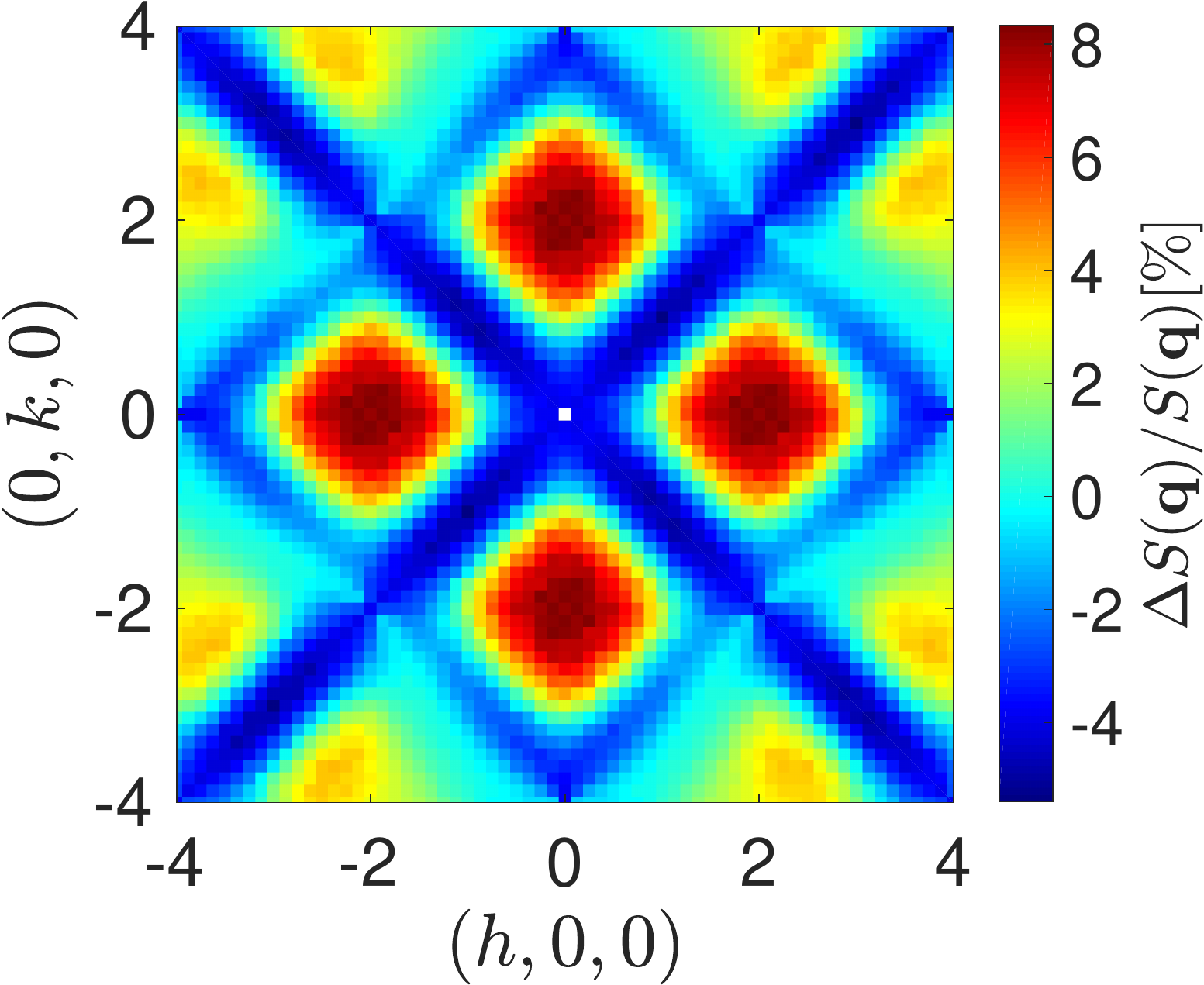}
	    \caption{$[001]$ SF.}\label{[001]SF}
	\end{subfigure}
	\begin{subfigure}[b]{0.22\textwidth}
	    \includegraphics[width=\textwidth]{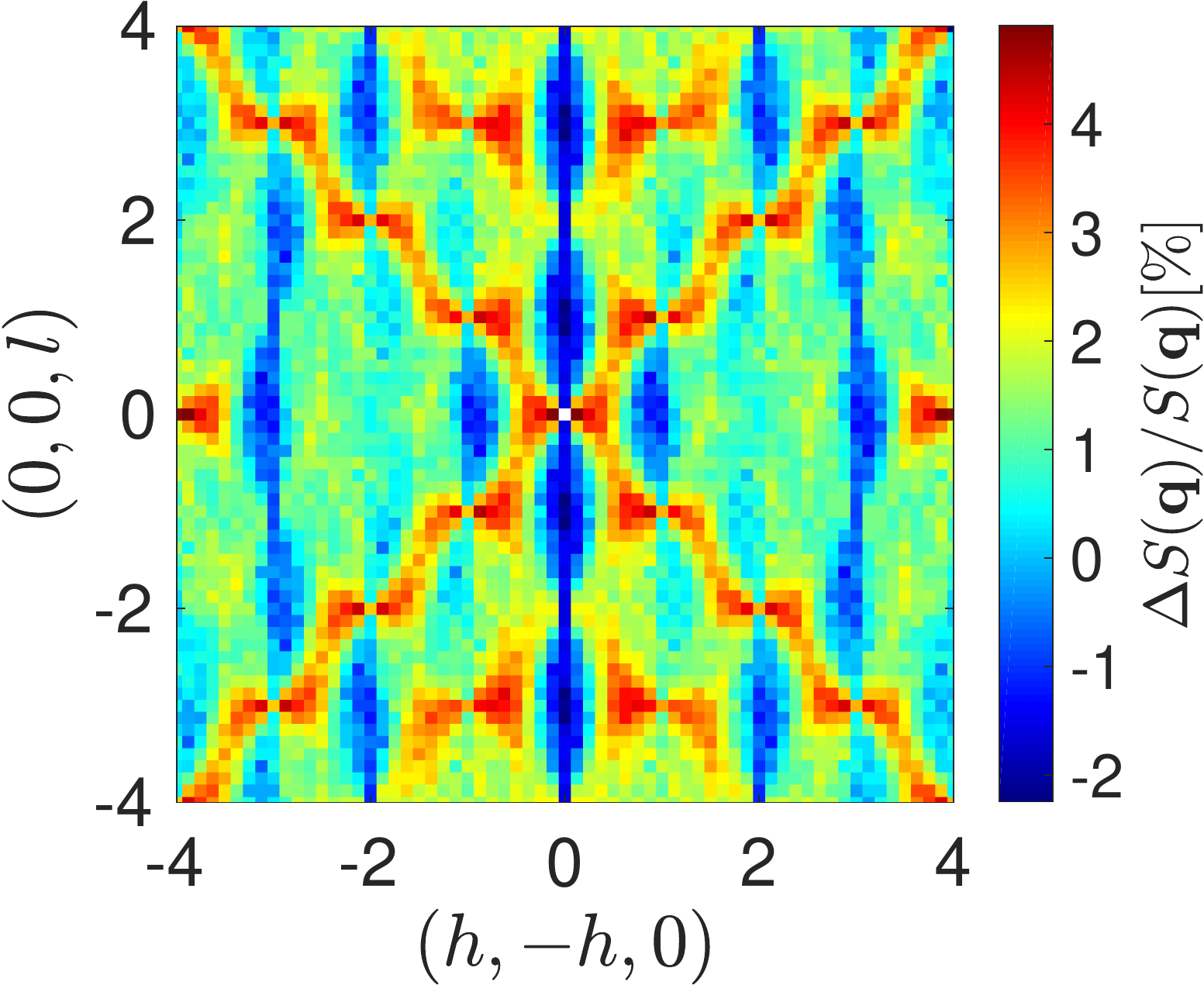}
	    \caption{$[110]$ Magnetic.}\label{[110]Total}
	\end{subfigure}
	\begin{subfigure}[b]{0.22\textwidth}
	    \includegraphics[width=\textwidth]{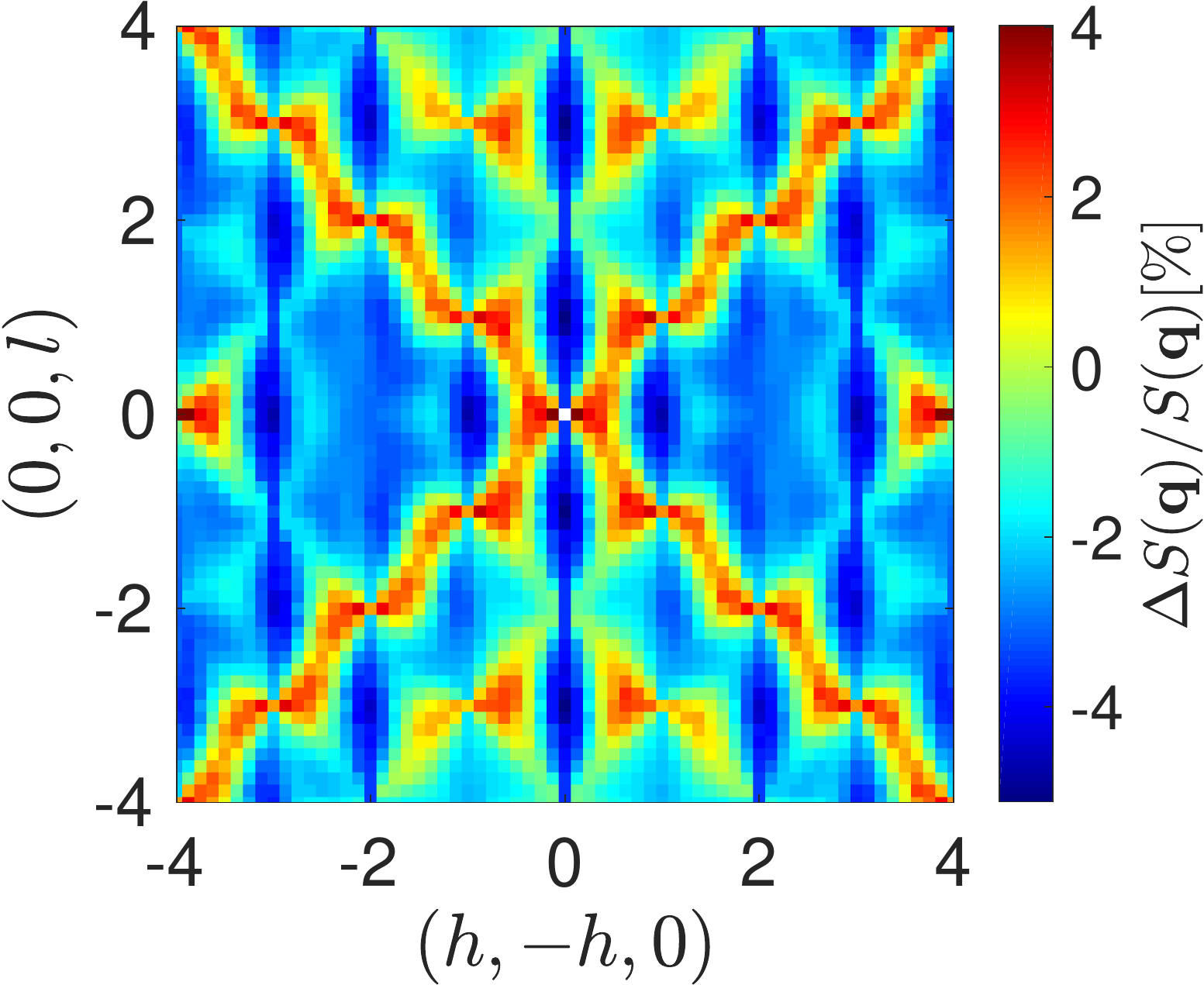}
	    \caption{$[110]$ SF.}\label{[110]SF}
	\end{subfigure}
	\begin{subfigure}[b]{0.22\textwidth}
	    \includegraphics[width=\textwidth]{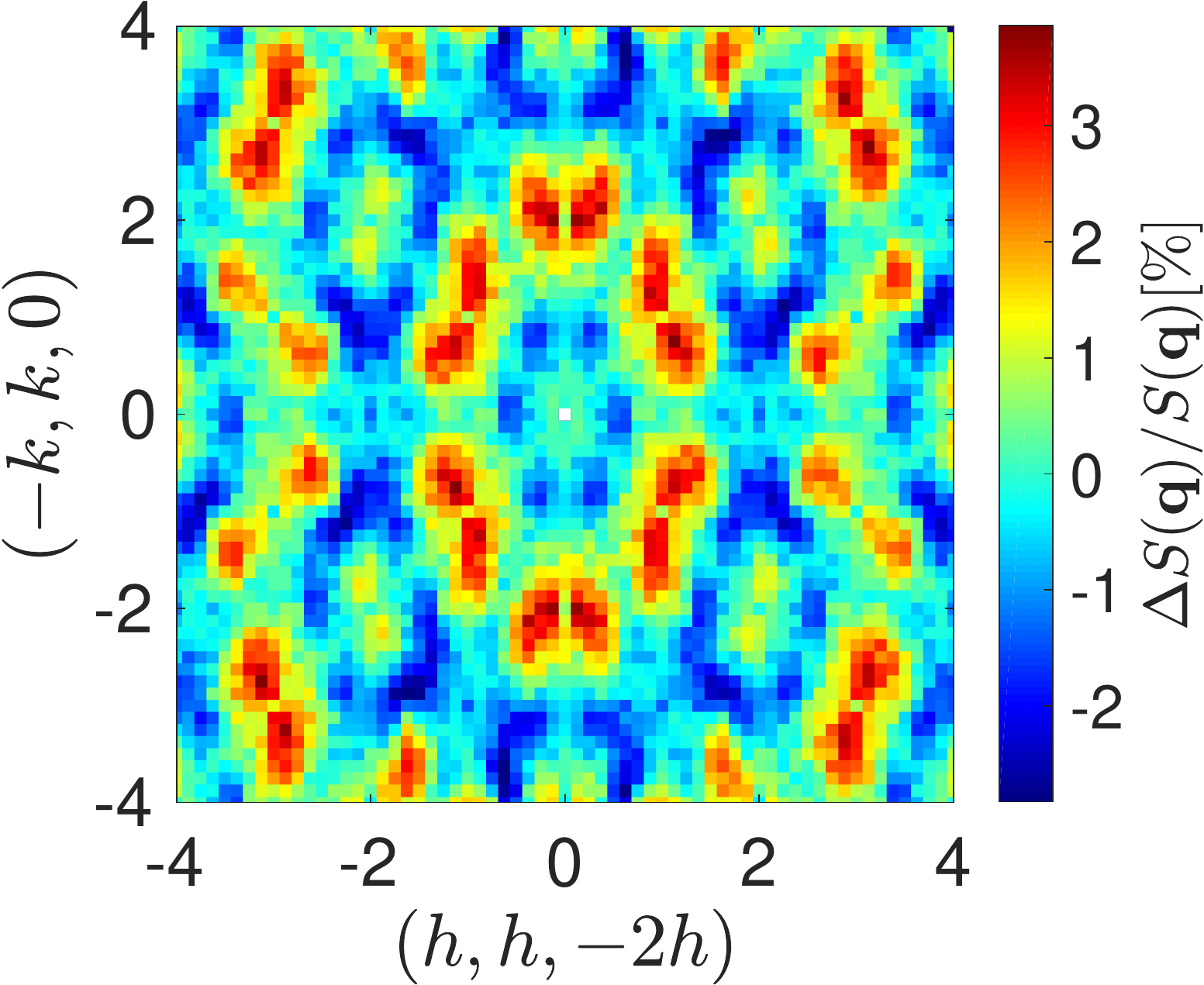}
	    \caption{$[111]$ Magnetic.}\label{[111]Total}
	\end{subfigure}
	\begin{subfigure}[b]{0.22\textwidth}
	    \includegraphics[width=\textwidth]{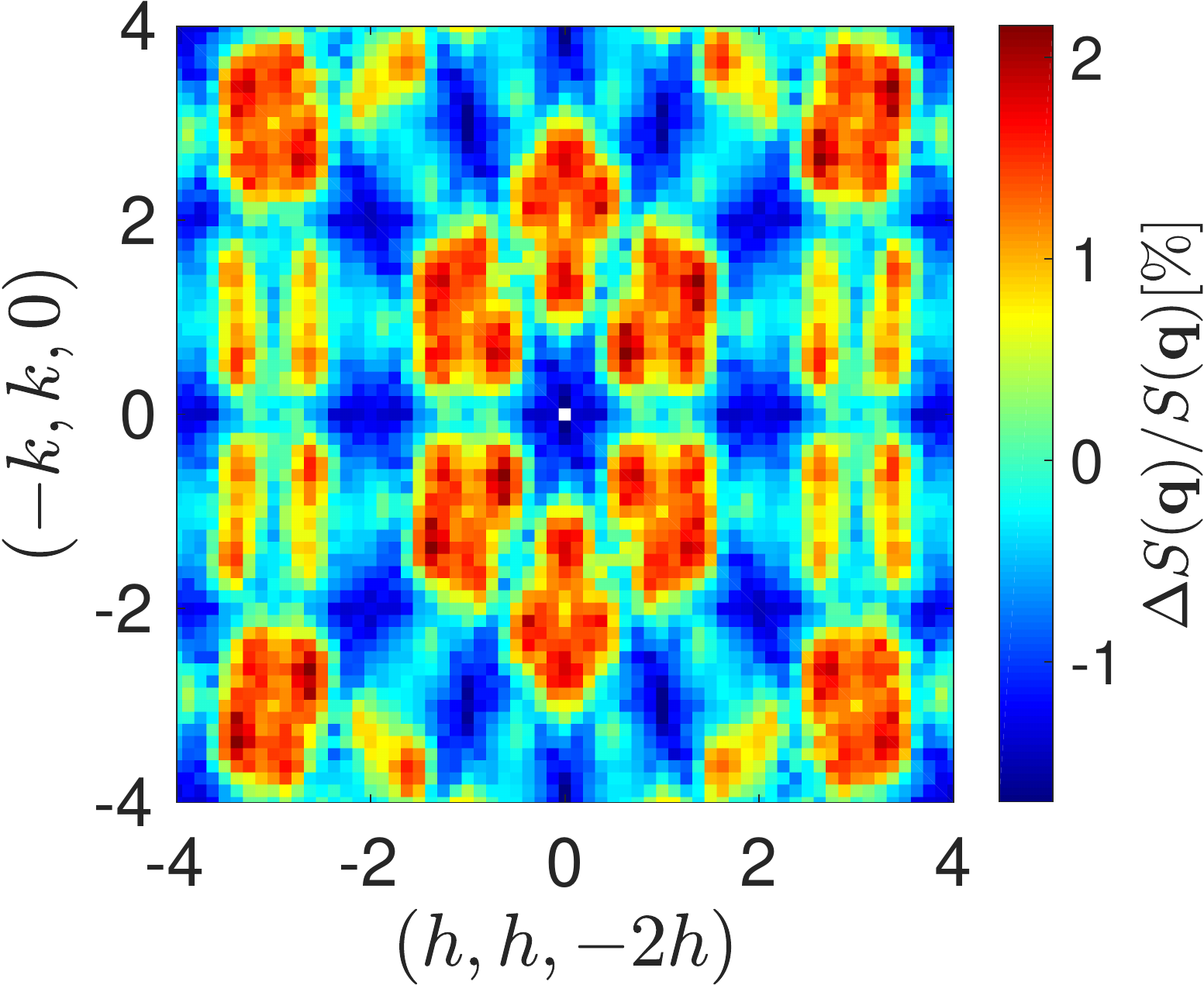}
	    \caption{$[111]$ SF.}\label{[111]SF}
	\end{subfigure}
	\caption{\raggedright 
	Predictions for the relative difference in the total magnetic scattering and the spin flip scattering by MC simulation on 8192 spins.  Exchange parameters and lattice compression are taken from the fit to magnetization measurements on DTO Fig.~\ref{fig:dM/Mcurves}. In all cases, the reciprocal plane is perpendicular to the direction of pressure and polarization is parallel with the direction of pressure.
	(a),(b) pressure along $[001]$ ($1.05$ GPa), (c),(d) pressure along $[110]$ ($1.20$ GPa), and (e),(f) pressure along $[111]$ ($1.30$ GPa), all at zero field and $T=1.7 \text{ K}$. 
	 }\label{fig:dsq}
\end{figure}

Based on the model parameters determined from the magnetization measurements in the previous section, our prediction for the relative increases in $S(\textbf{Q})$, $\Delta S(\textbf{q})/S(\textbf{q})=(S_{P>0}(\textbf{q})-S_{P=0}(\textbf{q}))/S_{P=0}(\textbf{q})$, at $T=1.7 \textup{ K}$ and pressures, $P_{001}=1.05 \text{ GPa}$, $P_{110}=1.20 \text{ GPa}$, $P_{111}=1.30 \text{ GPa}$ relevant for DTO are shown in Fig.~\ref{fig:dsq}. We find significant variations in the relative scattering intensities for the different scattering planes both for the magnetic and polarized spin flip cross sections. The largest variation can be seen for pressure along the $[001]$ direction with a relative increase of about $8\%$ in the $[001]$ spin flip channel of the $(h,k,0)$ plane, found for scattering vectors close to (2,0,0) and symmetry related points in reciprocal space.

\subsection{Neutron Diffraction Experiment on HTO}\label{experimentalSection}

In order to make a close comparison with the theoretical predictions of Fig.~\ref{fig:dsq} we would ideally like to perform neutron scattering measurements on  DTO crystals. However, the high neutron absorption cross section of natural dysprosium, cost of isotopically enriched samples, and the concurrent high probability of crystals cracking during the application of uniaxial pressure make these experiments inordinately expensive and challenging. Instead, we have chosen to perform neutron scattering measurements on crystals of HTO, which share many low temperature properties with DTO\cite{Springer_frust_book}.

Motivated by the theoretical results, uniaxial pressure was applied along the $[001]$ crystalline axis with $[001]$-polarized neutrons used to probe the $(h,k,0)$ plane. The axis perpendicular to the applied pressure is left unconstrained and therefore can give rise to the Poisson effect, in contrast to the previous magnetization measurements\cite{MITO}. However, within the $Q$-resolution of the instrument \cite{D7resolution}, no indicative change in the lattice parameters perpendicular to the pressure could be observed.

\begin{figure}[!h]
	\centering
	\includegraphics[width=1\linewidth]{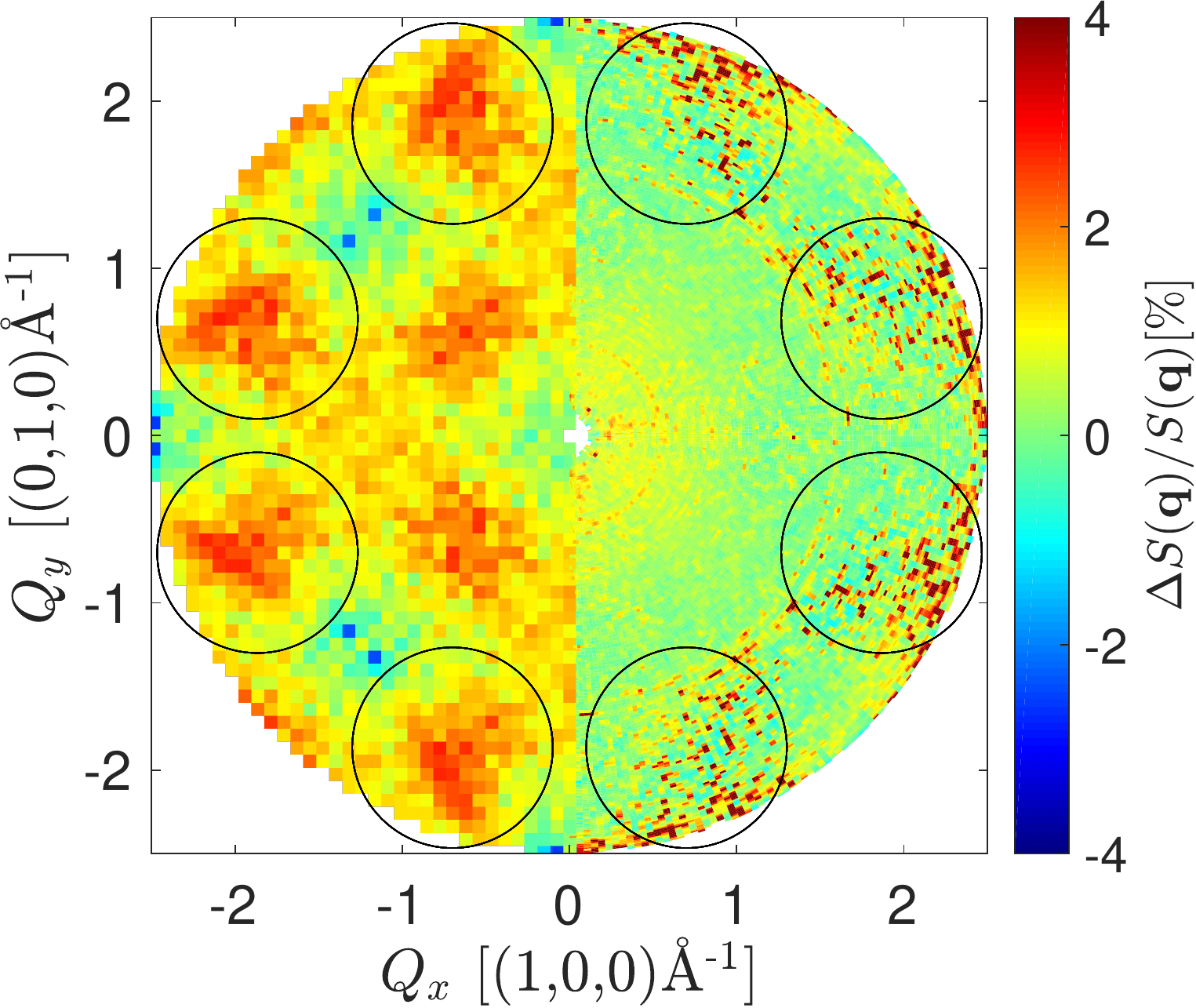}
	\caption{\raggedright Relative increases in the $[001]$ spin flip channel when pressure is applied along the $[001]$ direction.
	Left half: Estimate based on MC simulations of the model for 8192 spins ($J_\perp=1.56\textup{ K}$, $J_\parallel=1.975\textup{ K}$, $\kappa=2.8\%$). Right half: Experimental neutron scattering data after cell correction. Data taken at $1.5 \textup{ K}$ and $2.2\textup{ GPa}$. The circles drawn mark regions of increased intensity for high $Q$ in the experimental data and are plotted in symmetry equivalent places for the theoretical estimate.}\label{HTONS}
\end{figure}

We construct a theoretical estimate of the change in scattering intensity. Since we do not know the  pressure dependent susceptibility or magnetization data for HTO we are not able to make a detailed determination of the evolution of $J_\parallel$ and $\kappa$ with pressure in HTO (in contrast to DTO). However, since the two materials are chemically similar, and in order to make a rough estimate, we assume that the compression is linear in the applied pressure and that their compressibilities are identical. Using the fitted value for DTO ($\kappa=1.34\%$ at $1.05\textup{ GPa}$) then gives a compression of $\kappa=2.8\%$ for HTO at $2.2\textup{ GPa}$. The magnetic ions in HTO and DTO are significantly different, and it would be unreasonable to use the similarity argument also for the exchange interaction. Instead we adjust the single parameter $J_\parallel$ for HTO to give the best possible match with experiment.

A comparison between the theoretical prediction and the experimental results for the relative increase in the magnetic structure factor under $2.2\textup{ GPa}$ pressure, is shown in Fig.~\ref{HTONS}. A fourfold rotational averaging, consistent with the crystalline symmetry, has been performed on the experimental data. The experimental data have been corrected for a strong background contribution from the pressure cell \cite{cell}. This background dominates the low angular region of reciprocal space, leaving this region rather poorly sampled. 

There is a reasonable correspondence between the measurement and the theoretical prediction in the  region corresponding to wave-vector transfers $Q > 1.5 \textup{ \r{A}}^{-1}$, where the high-intensity regions are marked by circles. Further experimental activity will be required to improve the background from the pressure cell and reliably access the $Q < 1.5 \textup{ \r{A}}^{-1}$ region. In particular it is of interest to gather more statistics and to improve the construction and geometry of the cell to get a cleaner signal and better control of the pressure. The measurement of further crystallographic directions would also be of great value.

\begin{figure}[!h]
    \centering
    \begin{subfigure}[b]{0.22\textwidth}
	    \includegraphics[width=\textwidth]{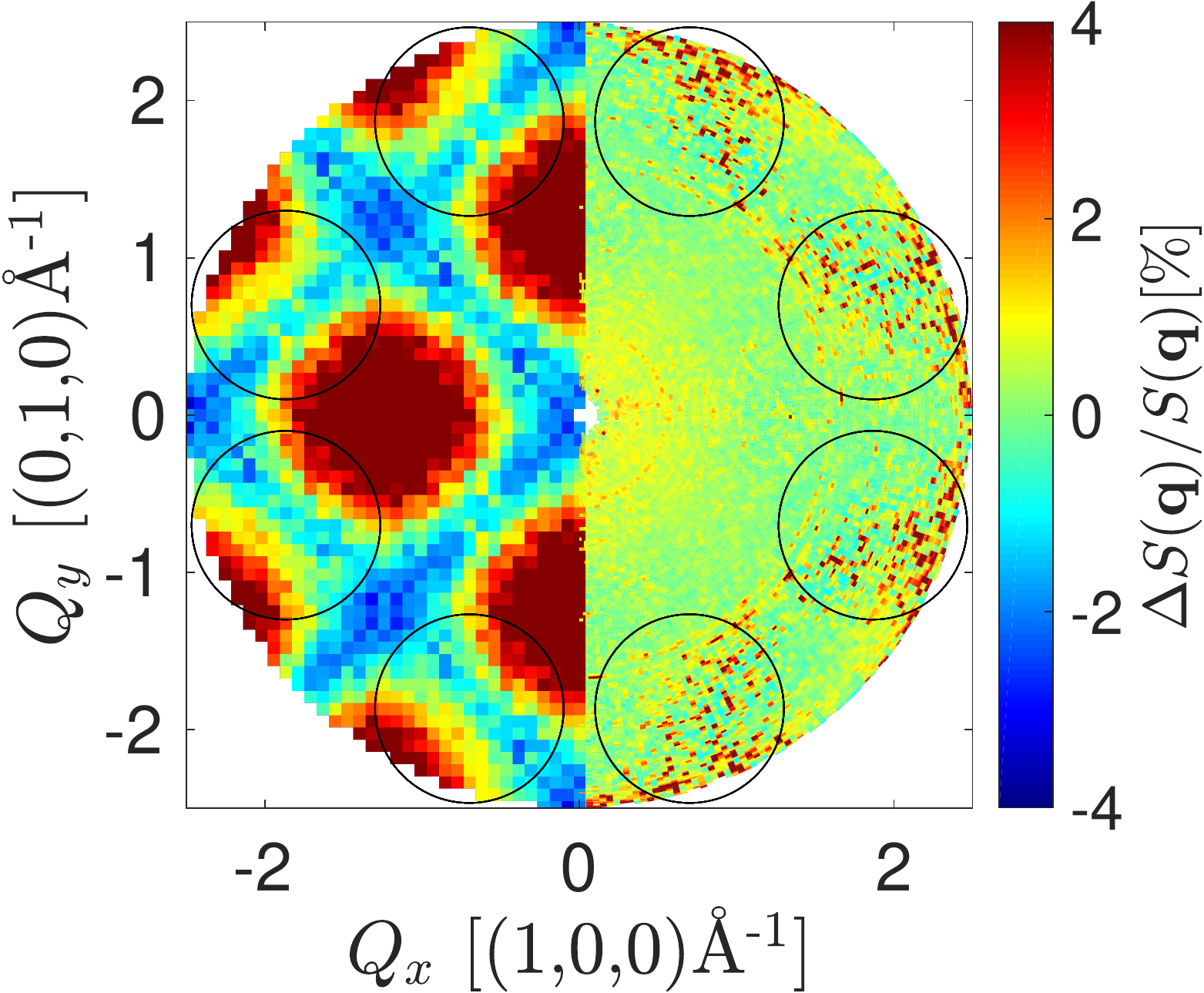}
	    \caption{$J_\parallel=1.94\textup{ K}$, $\kappa= 2.8\%$.}\label{SQ194}
	\end{subfigure}
	\begin{subfigure}[b]{0.22\textwidth}
	    \includegraphics[width=\textwidth]{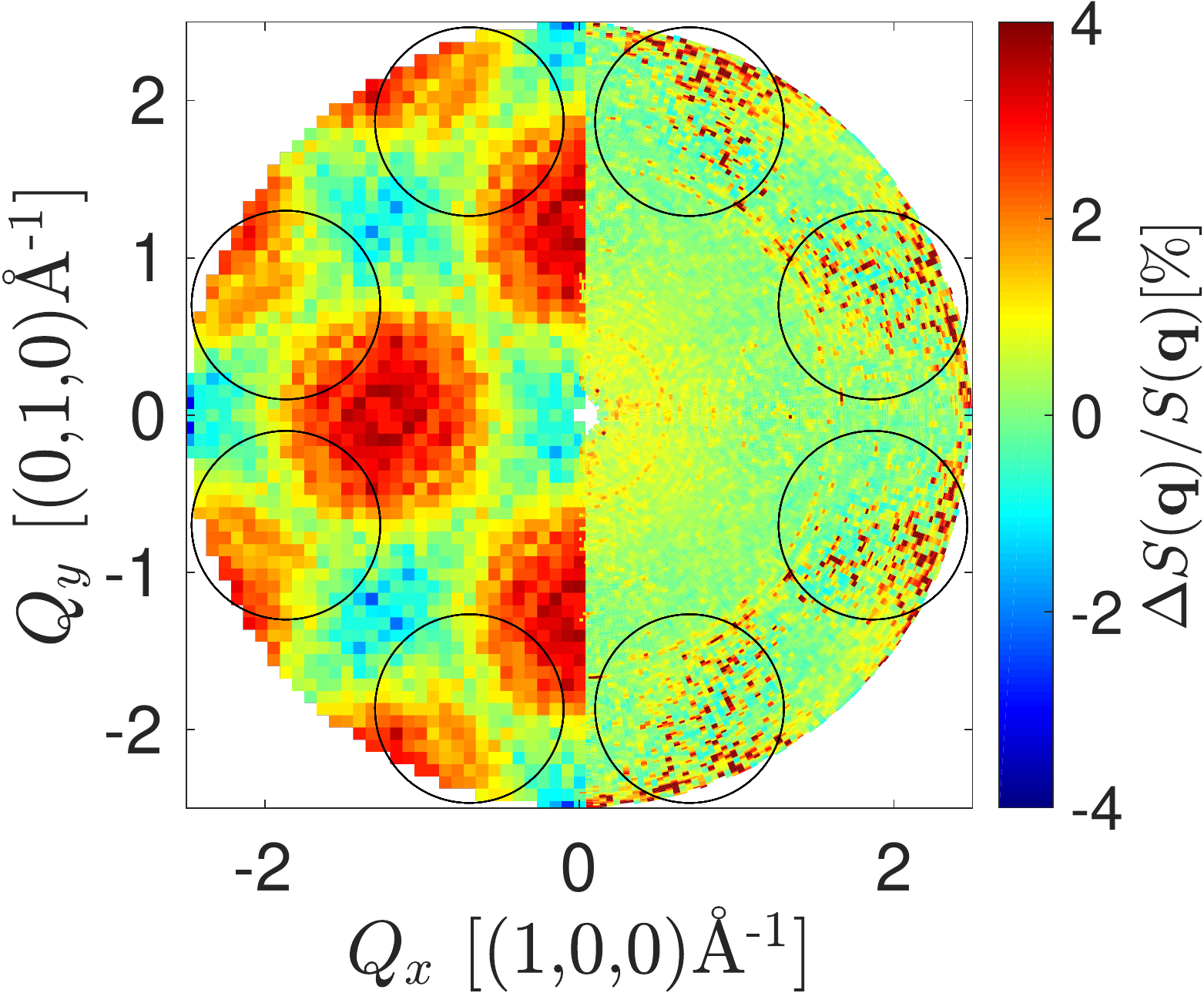}
	    \caption{$J_\parallel=1.96\textup{ K}$, $\kappa= 2.8\%$.}\label{SQ196}
	\end{subfigure}
	\begin{subfigure}[b]{0.22\textwidth}
	    \includegraphics[width=\textwidth]{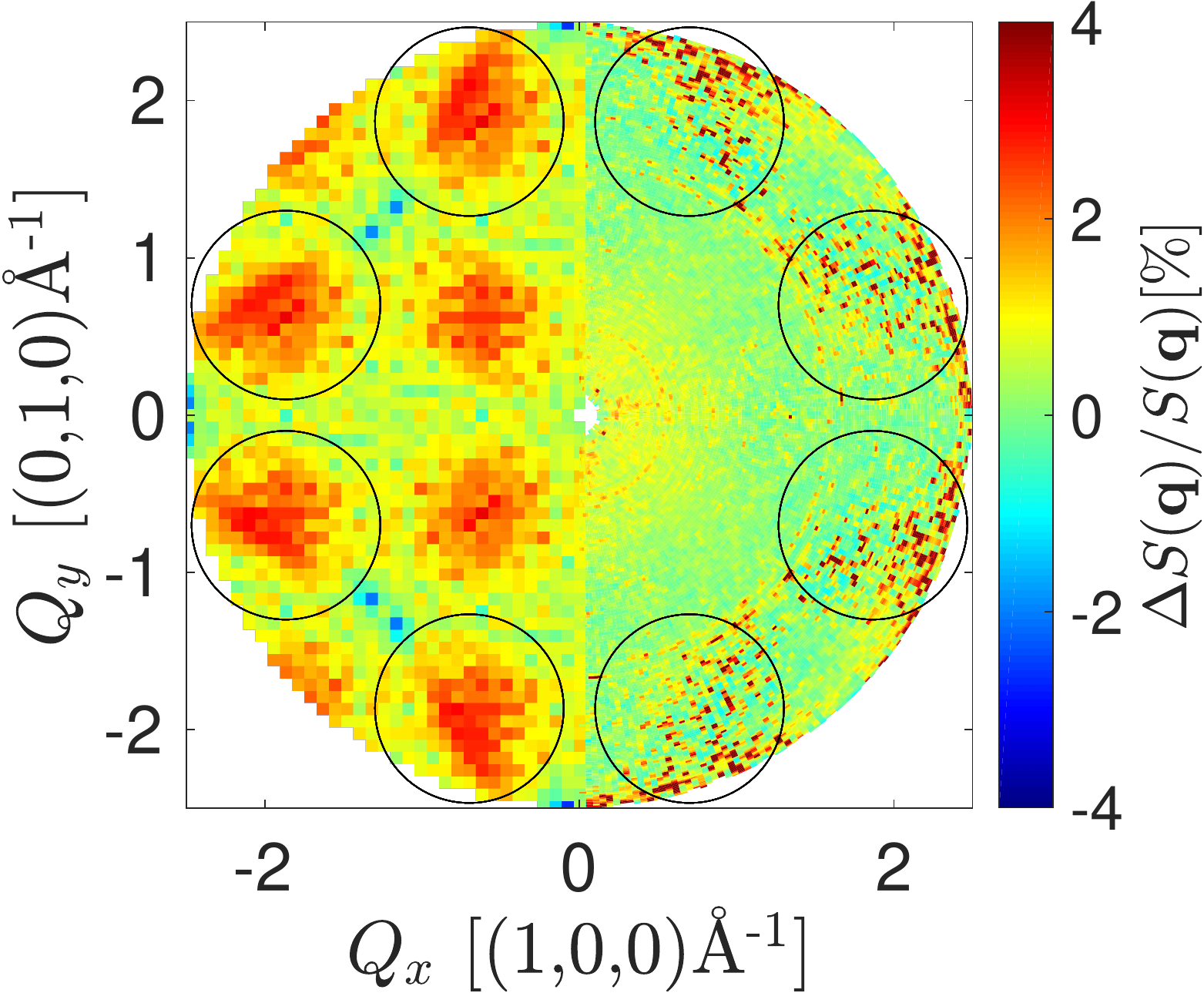}
	    \caption{$J_\parallel=1.98\textup{ K}$, $\kappa= 2.8\%$.}\label{SQ198}
	\end{subfigure}
	\begin{subfigure}[b]{0.22\textwidth}
	    \includegraphics[width=\textwidth]{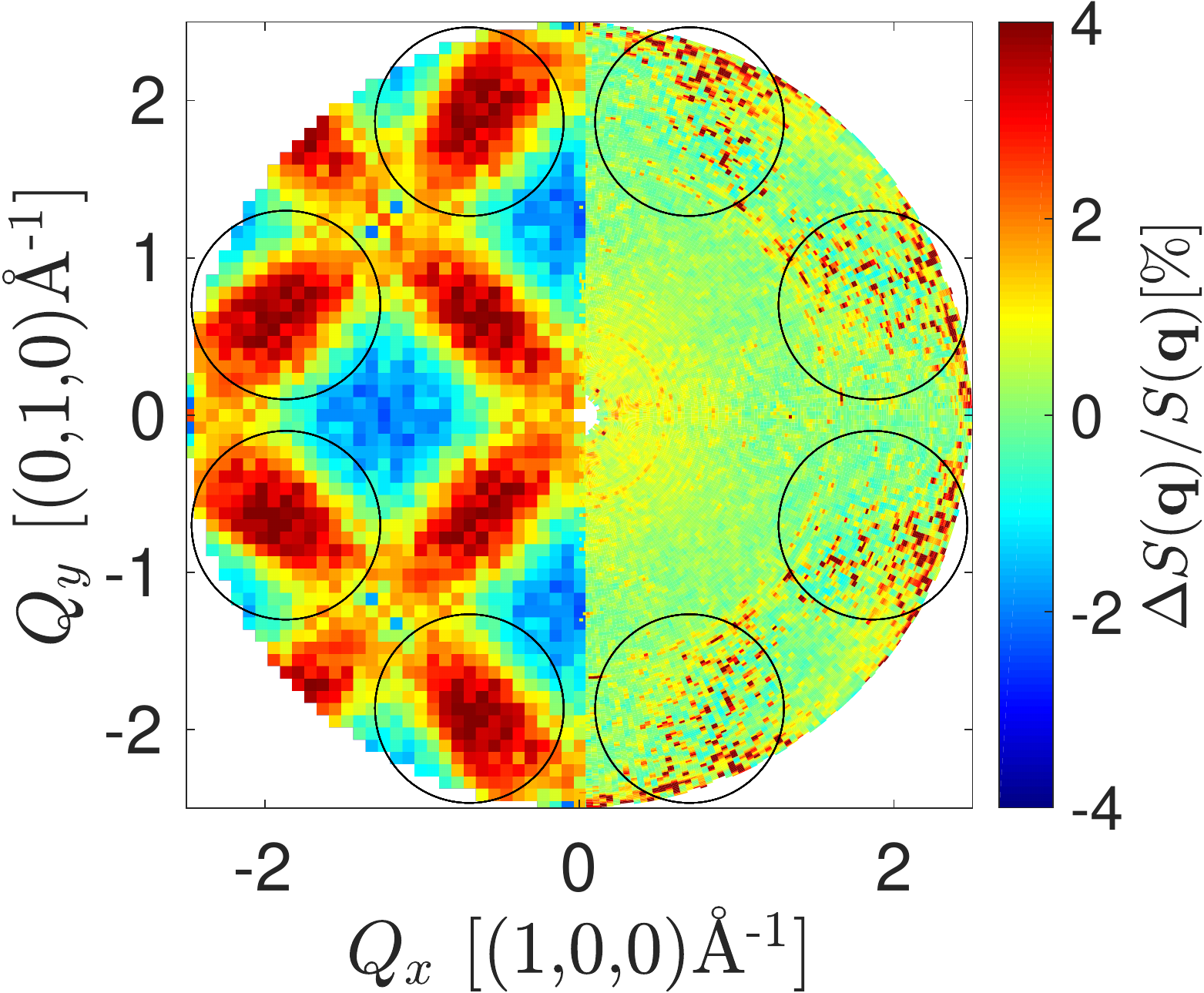}
	    \caption{$J_\parallel=2.00\textup{ K}$, $\kappa= 2.8\%$.}\label{SQ200}
	\end{subfigure}
    \caption{\raggedright Evolution of theoretical predictions for the relative increases in the $[001]$ spin flip channel when pressure is applied along the $[001]$ direction. Left half: MC simulation of the model for 8192 spins with varying $J_\parallel$ and $\kappa$ fixed. Right half: Experimental data (constant). We see that the cross/square pattern is inverted when we increase $J_\parallel$ from $1.94 \textup{ K}$ to $2.00\textup{ K}$. The circles drawn mark regions of increased intensity for high $Q$ in the experimental data and are plotted in symmetry equivalent places for the theoretical estimate.}\label{SQvsJplots}
\end{figure}

We note that the data from HTO has weaker features than those predicted for DTO, Fig.~\ref{[001]SF}. This can be accounted for by the tuning of the $J_\parallel$ parameter. Both changes in $\kappa$ and changes in $J_\parallel$ give rise to a cross/square pattern in the relative increases in the scattering, which are not prominent in the experimental data. The $J_\parallel$ parameter can be tuned such that these features are cancelled to a large degree. We choose it to get a profile as similar to the experiment as possible. Fig.~\ref{SQvsJplots} shows the evolution of the relative increases in the scattering (left half) when varying $J_\parallel$, together with the experimental data (right half). Note that the intensity in the cross/square pattern inverts as we increase $J_\parallel$ by only $3\%$, from $1.96\textup{ K}$ to $2.00\textup{ K}$. 

We determine that the closest fit has $J_\parallel=1.975 \sim 1.98 \textup{ K}$, about $27\%$ larger than $J_\perp$, in contrast to $J_\parallel/J_\perp-1=5\%$ seen for DTO. The fact that the experimental profile does not have a strong cross/square pattern implies, within our model, that the ground state of HTO lies close to the border between the $M_\perp$ and $M_\parallel$ states. This is because the inversion point of the cross/square pattern is at the border between the $M_\parallel$ and $M_\perp$ phases in the ground state phase diagram, Fig.~\ref{fig:phaseDiagram}.

\section{Discussion and Conclusion}

Using a dipolar spin ice model, we have accurately modeled the experimental changes in magnetization of DTO under the application of pressure in multiple crystalline directions. Using the model parameters derived from the magnetization measurements we have calculated the correspondingly expected changes in neutron scattering cross section for DTO. Neutron scattering experiments were performed on HTO under applied pressure, and we found the response to be weaker than what we predicted for DTO. One possible reason for the difference is that HTO may be located at the boundary between two different chain states, as discussed in the previous section.  Another is that  the compressibility of HTO may be smaller than that of DTO, and that the associated pressure induced changes in the exchange parameters are also smaller. Furthermore, we have not considered changes to exchange parameters past the nearest neighbors, which certainly could influence the results \cite{yavo08,henelius16}. In conclusion, we note that in order to determine which of these scenarios are relevant for HTO, more experiments are necessary. In particular similar  magnetization measurements of HTO under pressure, as previously were performed on DTO, would be highly useful.

\acknowledgements
We thank M. Mito and K. Matsuhira for useful discussions and sharing their data. The neutron scattering experiments were performed at the Paul Scherrer Institute and the Institute Laue-Langevin. The project was supported by Nordforsk through the program NNSP and by the Danish Agency for Research and Innovation through DANSCATT.

%\bibliography{references} 

\begin{thebibliography}{28}%
	\makeatletter
	\providecommand \@ifxundefined [1]{%
		\@ifx{#1\undefined}
	}%
	\providecommand \@ifnum [1]{%
		\ifnum #1\expandafter \@firstoftwo
		\else \expandafter \@secondoftwo
		\fi
	}%
	\providecommand \@ifx [1]{%
		\ifx #1\expandafter \@firstoftwo
		\else \expandafter \@secondoftwo
		\fi
	}%
	\providecommand \natexlab [1]{#1}%
	\providecommand \enquote  [1]{``#1''}%
	\providecommand \bibnamefont  [1]{#1}%
	\providecommand \bibfnamefont [1]{#1}%
	\providecommand \citenamefont [1]{#1}%
	\providecommand \href@noop [0]{\@secondoftwo}%
	\providecommand \href [0]{\begingroup \@sanitize@url \@href}%
	\providecommand \@href[1]{\@@startlink{#1}\@@href}%
	\providecommand \@@href[1]{\endgroup#1\@@endlink}%
	\providecommand \@sanitize@url [0]{\catcode `\\12\catcode `\$12\catcode
		`\&12\catcode `\#12\catcode `\^12\catcode `\_12\catcode `\%12\relax}%
	\providecommand \@@startlink[1]{}%
	\providecommand \@@endlink[0]{}%
	\providecommand \url  [0]{\begingroup\@sanitize@url \@url }%
	\providecommand \@url [1]{\endgroup\@href {#1}{\urlprefix }}%
	\providecommand \urlprefix  [0]{URL }%
	\providecommand \Eprint [0]{\href }%
	\providecommand \doibase [0]{http://dx.doi.org/}%
	\providecommand \selectlanguage [0]{\@gobble}%
	\providecommand \bibinfo  [0]{\@secondoftwo}%
	\providecommand \bibfield  [0]{\@secondoftwo}%
	\providecommand \translation [1]{[#1]}%
	\providecommand \BibitemOpen [0]{}%
	\providecommand \bibitemStop [0]{}%
	\providecommand \bibitemNoStop [0]{.\EOS\space}%
	\providecommand \EOS [0]{\spacefactor3000\relax}%
	\providecommand \BibitemShut  [1]{\csname bibitem#1\endcsname}%
	\let\auto@bib@innerbib\@empty
	%</preamble>
	\bibitem [{\citenamefont {Lacroix}\ \emph {et~al.}(2011)\citenamefont
		{Lacroix}, \citenamefont {Mendels},\ and\ \citenamefont
		{Mila}}]{Springer_frust_book}%
	\BibitemOpen
	\bibinfo {editor} {\bibfnamefont {C.}~\bibnamefont {Lacroix}}, \bibinfo
	{editor} {\bibfnamefont {P.}~\bibnamefont {Mendels}}, \ and\ \bibinfo
	{editor} {\bibfnamefont {F.}~\bibnamefont {Mila}},\ eds.,\ \href@noop {}
	{\emph {\bibinfo {title} {Highly Frustrated Magnetism}}},\ \bibinfo {series}
	{Springer Series in Solid-State Sciences}, Vol.\ \bibinfo {volume} {164}\
	(\bibinfo  {publisher} {Springer},\ \bibinfo {year} {2011})\BibitemShut
	{NoStop}%
	\bibitem [{\citenamefont {Castelnovo}\ \emph {et~al.}(2008)\citenamefont
		{Castelnovo}, \citenamefont {Moessner},\ and\ \citenamefont
		{Sondhi}}]{Castelnovo_Nature}%
	\BibitemOpen
	\bibfield  {author} {\bibinfo {author} {\bibfnamefont {C.}~\bibnamefont
			{Castelnovo}}, \bibinfo {author} {\bibfnamefont {R.}~\bibnamefont
			{Moessner}}, \ and\ \bibinfo {author} {\bibfnamefont {S.~L.}\ \bibnamefont
			{Sondhi}},\ }\href@noop {} {\bibfield  {journal} {\bibinfo  {journal}
			{Nature}\ }\textbf {\bibinfo {volume} {451}},\ \bibinfo {pages} {42}
		(\bibinfo {year} {2008})}\BibitemShut {NoStop}%
	\bibitem [{\citenamefont {Brooks-Bartlett}\ \emph {et~al.}(2014)\citenamefont
		{Brooks-Bartlett}, \citenamefont {Banks}, \citenamefont {Jaubert},
		\citenamefont {Harman-Clarke},\ and\ \citenamefont {Holdsworth}}]{Brooks}%
	\BibitemOpen
	\bibfield  {author} {\bibinfo {author} {\bibfnamefont {M.~E.}\ \bibnamefont
			{Brooks-Bartlett}}, \bibinfo {author} {\bibfnamefont {S.~T.}\ \bibnamefont
			{Banks}}, \bibinfo {author} {\bibfnamefont {L.~D.~C.}\ \bibnamefont
			{Jaubert}}, \bibinfo {author} {\bibfnamefont {A.}~\bibnamefont
			{Harman-Clarke}}, \ and\ \bibinfo {author} {\bibfnamefont {P.~C.~W.}\
			\bibnamefont {Holdsworth}},\ }\href {\doibase 10.1103/PhysRevX.4.011007}
	{\bibfield  {journal} {\bibinfo  {journal} {Phys. Rev. X}\ }\textbf {\bibinfo
			{volume} {4}},\ \bibinfo {pages} {011007} (\bibinfo {year}
		{2014})}\BibitemShut {NoStop}%
	\bibitem [{\citenamefont {Henley}(2005)}]{Henley_Coulomb}%
	\BibitemOpen
	\bibfield  {author} {\bibinfo {author} {\bibfnamefont {C.~L.}\ \bibnamefont
			{Henley}},\ }\href@noop {} {\bibfield  {journal} {\bibinfo  {journal} {Phys.
				Rev. B}\ }\textbf {\bibinfo {volume} {71}},\ \bibinfo {pages} {014424}
		(\bibinfo {year} {2005})}\BibitemShut {NoStop}%
	\bibitem [{\citenamefont {Pomaranski}\ \emph {et~al.}(2013)\citenamefont
		{Pomaranski}, \citenamefont {Yaraskavitch}, \citenamefont {Meng},
		\citenamefont {Ross}, \citenamefont {Noad}, \citenamefont {Dabkowska},
		\citenamefont {Gaulin},\ and\ \citenamefont {Kycia}}]{Poma13}%
	\BibitemOpen
	\bibfield  {author} {\bibinfo {author} {\bibfnamefont {D.}~\bibnamefont
			{Pomaranski}}, \bibinfo {author} {\bibfnamefont {L.~R.}\ \bibnamefont
			{Yaraskavitch}}, \bibinfo {author} {\bibfnamefont {S.}~\bibnamefont {Meng}},
		\bibinfo {author} {\bibfnamefont {K.~A.}\ \bibnamefont {Ross}}, \bibinfo
		{author} {\bibfnamefont {H.~M.~L.}\ \bibnamefont {Noad}}, \bibinfo {author}
		{\bibfnamefont {H.~A.}\ \bibnamefont {Dabkowska}}, \bibinfo {author}
		{\bibfnamefont {B.~D.}\ \bibnamefont {Gaulin}}, \ and\ \bibinfo {author}
		{\bibfnamefont {J.~B.}\ \bibnamefont {Kycia}},\ }\href@noop {} {\bibfield
		{journal} {\bibinfo  {journal} {{Nature Physics}}\ }\textbf {\bibinfo
			{volume} {{9}}},\ \bibinfo {pages} {{353}} (\bibinfo {year}
		{{2013}})}\BibitemShut {NoStop}%
	\bibitem [{\citenamefont {Henelius}\ \emph {et~al.}(2016)\citenamefont
		{Henelius}, \citenamefont {Lin}, \citenamefont {Enjalran}, \citenamefont
		{Hao}, \citenamefont {Rau}, \citenamefont {Altosaar}, \citenamefont
		{Flicker}, \citenamefont {Yavors'kii},\ and\ \citenamefont
		{Gingras}}]{henelius16}%
	\BibitemOpen
	\bibfield  {author} {\bibinfo {author} {\bibfnamefont {P.}~\bibnamefont
			{Henelius}}, \bibinfo {author} {\bibfnamefont {T.}~\bibnamefont {Lin}},
		\bibinfo {author} {\bibfnamefont {M.}~\bibnamefont {Enjalran}}, \bibinfo
		{author} {\bibfnamefont {Z.}~\bibnamefont {Hao}}, \bibinfo {author}
		{\bibfnamefont {J.~G.}\ \bibnamefont {Rau}}, \bibinfo {author} {\bibfnamefont
			{J.}~\bibnamefont {Altosaar}}, \bibinfo {author} {\bibfnamefont
			{F.}~\bibnamefont {Flicker}}, \bibinfo {author} {\bibfnamefont
			{T.}~\bibnamefont {Yavors'kii}}, \ and\ \bibinfo {author} {\bibfnamefont
			{M.~J.~P.}\ \bibnamefont {Gingras}},\ }\href {\doibase
		10.1103/PhysRevB.93.024402} {\bibfield  {journal} {\bibinfo  {journal} {Phys.
				Rev. B}\ }\textbf {\bibinfo {volume} {93}},\ \bibinfo {pages} {024402}
		(\bibinfo {year} {2016})}\BibitemShut {NoStop}%
	\bibitem [{\citenamefont {Melko}\ and\ \citenamefont
		{Gingras}(2004)}]{MelkoGingras}%
	\BibitemOpen
	\bibfield  {author} {\bibinfo {author} {\bibfnamefont {R.~G.}\ \bibnamefont
			{Melko}}\ and\ \bibinfo {author} {\bibfnamefont {M.~J.~P.}\ \bibnamefont
			{Gingras}},\ }\href {http://stacks.iop.org/0953-8984/16/i=43/a=R02}
	{\bibfield  {journal} {\bibinfo  {journal} {Journal of Physics: Condensed
				Matter}\ }\textbf {\bibinfo {volume} {16}},\ \bibinfo {pages} {R1277}
		(\bibinfo {year} {2004})}\BibitemShut {NoStop}%
	\bibitem [{\citenamefont {Mito}\ \emph {et~al.}(2007)\citenamefont {Mito},
		\citenamefont {Kuwabara}, \citenamefont {Matsuhira}, \citenamefont {Deguchi},
		\citenamefont {Takagi},\ and\ \citenamefont {Hiroi}}]{MITO}%
	\BibitemOpen
	\bibfield  {author} {\bibinfo {author} {\bibfnamefont {M.}~\bibnamefont
			{Mito}}, \bibinfo {author} {\bibfnamefont {S.}~\bibnamefont {Kuwabara}},
		\bibinfo {author} {\bibfnamefont {K.}~\bibnamefont {Matsuhira}}, \bibinfo
		{author} {\bibfnamefont {H.}~\bibnamefont {Deguchi}}, \bibinfo {author}
		{\bibfnamefont {S.}~\bibnamefont {Takagi}}, \ and\ \bibinfo {author}
		{\bibfnamefont {Z.}~\bibnamefont {Hiroi}},\ }\href {\doibase
		https://doi.org/10.1016/j.jmmm.2006.10.441} {\bibfield  {journal} {\bibinfo
			{journal} {Journal of Magnetism and Magnetic Materials}\ }\textbf {\bibinfo
			{volume} {310}},\ \bibinfo {pages} {e432 } (\bibinfo {year} {2007})},\
	\bibinfo {note} {proceedings of the 17th International Conference on
		Magnetism}\BibitemShut {NoStop}%
	\bibitem [{\citenamefont {Mirebeau}\ and\ \citenamefont
		{Goncharenko}(2004)}]{Mirebeau_2004}%
	\BibitemOpen
	\bibfield  {author} {\bibinfo {author} {\bibfnamefont {I.}~\bibnamefont
			{Mirebeau}}\ and\ \bibinfo {author} {\bibfnamefont {I.}~\bibnamefont
			{Goncharenko}},\ }\href {\doibase 10.1088/0953-8984/16/11/012} {\bibfield
		{journal} {\bibinfo  {journal} {Journal of Physics: Condensed Matter}\
		}\textbf {\bibinfo {volume} {16}},\ \bibinfo {pages} {S653} (\bibinfo {year}
		{2004})}\BibitemShut {NoStop}%
	\bibitem [{\citenamefont {Jaubert}\ \emph {et~al.}(2010)\citenamefont
		{Jaubert}, \citenamefont {Chalker}, \citenamefont {Holdsworth},\ and\
		\citenamefont {Moessner}}]{Jaubert1}%
	\BibitemOpen
	\bibfield  {author} {\bibinfo {author} {\bibfnamefont {L.~D.~C.}\
			\bibnamefont {Jaubert}}, \bibinfo {author} {\bibfnamefont {J.~T.}\
			\bibnamefont {Chalker}}, \bibinfo {author} {\bibfnamefont {P.~C.~W.}\
			\bibnamefont {Holdsworth}}, \ and\ \bibinfo {author} {\bibfnamefont
			{R.}~\bibnamefont {Moessner}},\ }\href {\doibase
		10.1103/PhysRevLett.105.087201} {\bibfield  {journal} {\bibinfo  {journal}
			{Phys. Rev. Lett.}\ }\textbf {\bibinfo {volume} {105}},\ \bibinfo {pages}
		{087201} (\bibinfo {year} {2010})}\BibitemShut {NoStop}%
	\bibitem [{\citenamefont {Bramwell}\ \emph {et~al.}(2009)\citenamefont
		{Bramwell}, \citenamefont {Giblin}, \citenamefont {Calder}, \citenamefont
		{Aldus}, \citenamefont {Prabhakaran},\ and\ \citenamefont
		{Fennell}}]{Bramwell_Nature}%
	\BibitemOpen
	\bibfield  {author} {\bibinfo {author} {\bibfnamefont {S.~T.}\ \bibnamefont
			{Bramwell}}, \bibinfo {author} {\bibfnamefont {S.~R.}\ \bibnamefont
			{Giblin}}, \bibinfo {author} {\bibfnamefont {S.}~\bibnamefont {Calder}},
		\bibinfo {author} {\bibfnamefont {R.}~\bibnamefont {Aldus}}, \bibinfo
		{author} {\bibfnamefont {D.}~\bibnamefont {Prabhakaran}}, \ and\ \bibinfo
		{author} {\bibfnamefont {T.}~\bibnamefont {Fennell}},\ }\href@noop {}
	{\bibfield  {journal} {\bibinfo  {journal} {Nature}\ }\textbf {\bibinfo
			{volume} {461}},\ \bibinfo {pages} {956} (\bibinfo {year}
		{2009})}\BibitemShut {NoStop}%
	\bibitem [{\citenamefont {den Hertog}\ and\ \citenamefont
		{Gingras}(2000{\natexlab{a}})}]{hertog00}%
	\BibitemOpen
	\bibfield  {author} {\bibinfo {author} {\bibfnamefont {B.~C.}\ \bibnamefont
			{den Hertog}}\ and\ \bibinfo {author} {\bibfnamefont {M.~J.~P.}\ \bibnamefont
			{Gingras}},\ }\href {\doibase 10.1103/PhysRevLett.84.3430} {\bibfield
		{journal} {\bibinfo  {journal} {Phys. Rev. Lett.}\ }\textbf {\bibinfo
			{volume} {84}},\ \bibinfo {pages} {3430} (\bibinfo {year}
		{2000}{\natexlab{a}})}\BibitemShut {NoStop}%
	\bibitem [{\citenamefont {Yavors'kii}\ \emph {et~al.}(2008)\citenamefont
		{Yavors'kii}, \citenamefont {Fennell}, \citenamefont {Gingras},\ and\
		\citenamefont {Bramwell}}]{yavo08}%
	\BibitemOpen
	\bibfield  {author} {\bibinfo {author} {\bibfnamefont {T.}~\bibnamefont
			{Yavors'kii}}, \bibinfo {author} {\bibfnamefont {T.}~\bibnamefont {Fennell}},
		\bibinfo {author} {\bibfnamefont {M.~J.~P.}\ \bibnamefont {Gingras}}, \ and\
		\bibinfo {author} {\bibfnamefont {S.~T.}\ \bibnamefont {Bramwell}},\
	}\href@noop {} {\bibfield  {journal} {\bibinfo  {journal} {Phys. Rev. Lett.}\
		}\textbf {\bibinfo {volume} {101}},\ \bibinfo {pages} {037204} (\bibinfo
		{year} {2008})}\BibitemShut {NoStop}%
	\bibitem [{\citenamefont {den Hertog}\ and\ \citenamefont
		{Gingras}(2000{\natexlab{b}})}]{denHertog}%
	\BibitemOpen
	\bibfield  {author} {\bibinfo {author} {\bibfnamefont {B.~C.}\ \bibnamefont
			{den Hertog}}\ and\ \bibinfo {author} {\bibfnamefont {M.~J.~P.}\ \bibnamefont
			{Gingras}},\ }\href@noop {} {\bibfield  {journal} {\bibinfo  {journal} {Phys.
				Rev. Lett.}\ }\textbf {\bibinfo {volume} {84}},\ \bibinfo {pages} {3430}
		(\bibinfo {year} {2000}{\natexlab{b}})}\BibitemShut {NoStop}%
	\bibitem [{\citenamefont {Isakov}\ \emph {et~al.}(2005)\citenamefont {Isakov},
		\citenamefont {Moessner},\ and\ \citenamefont {Sondhi}}]{Isakov_SS}%
	\BibitemOpen
	\bibfield  {author} {\bibinfo {author} {\bibfnamefont {S.~V.}\ \bibnamefont
			{Isakov}}, \bibinfo {author} {\bibfnamefont {R.}~\bibnamefont {Moessner}}, \
		and\ \bibinfo {author} {\bibfnamefont {S.~L.}\ \bibnamefont {Sondhi}},\
	}\href@noop {} {\bibfield  {journal} {\bibinfo  {journal} {Phys. Rev. Lett.}\
		}\textbf {\bibinfo {volume} {95}},\ \bibinfo {pages} {217201} (\bibinfo
		{year} {2005})}\BibitemShut {NoStop}%
	\bibitem [{\citenamefont {Jaubert}\ \emph {et~al.}(2013)\citenamefont
		{Jaubert}, \citenamefont {Harris}, \citenamefont {Fennell}, \citenamefont
		{Melko}, \citenamefont {Bramwell},\ and\ \citenamefont
		{Holdsworth}}]{jaubert13}%
	\BibitemOpen
	\bibfield  {author} {\bibinfo {author} {\bibfnamefont {L.~D.~C.}\
			\bibnamefont {Jaubert}}, \bibinfo {author} {\bibfnamefont {M.~J.}\
			\bibnamefont {Harris}}, \bibinfo {author} {\bibfnamefont {T.}~\bibnamefont
			{Fennell}}, \bibinfo {author} {\bibfnamefont {R.~G.}\ \bibnamefont {Melko}},
		\bibinfo {author} {\bibfnamefont {S.~T.}\ \bibnamefont {Bramwell}}, \ and\
		\bibinfo {author} {\bibfnamefont {P.~C.~W.}\ \bibnamefont {Holdsworth}},\
	}\href {\doibase 10.1103/PhysRevX.3.011014} {\bibfield  {journal} {\bibinfo
			{journal} {Phys. Rev. X}\ }\textbf {\bibinfo {volume} {3}},\ \bibinfo {pages}
		{011014} (\bibinfo {year} {2013})}\BibitemShut {NoStop}%
	\bibitem [{\citenamefont {Jaubert}\ and\ \citenamefont
		{Holdsworth}(2011)}]{Jaubert_2011}%
	\BibitemOpen
	\bibfield  {author} {\bibinfo {author} {\bibfnamefont {L.~D.~C.}\
			\bibnamefont {Jaubert}}\ and\ \bibinfo {author} {\bibfnamefont {P.~C.~W.}\
			\bibnamefont {Holdsworth}},\ }\href {\doibase 10.1088/0953-8984/23/16/164222}
	{\bibfield  {journal} {\bibinfo  {journal} {Journal of Physics: Condensed
				Matter}\ }\textbf {\bibinfo {volume} {23}},\ \bibinfo {pages} {164222}
		(\bibinfo {year} {2011})}\BibitemShut {NoStop}%
	\bibitem [{\citenamefont {Frenkel}\ and\ \citenamefont
		{Smit}(2001)}]{frenkel2001understanding}%
	\BibitemOpen
	\bibfield  {author} {\bibinfo {author} {\bibfnamefont {D.}~\bibnamefont
			{Frenkel}}\ and\ \bibinfo {author} {\bibfnamefont {B.}~\bibnamefont {Smit}},\
	}\href {https://books.google.se/books?id=5qTzldS9ROIC} {\emph {\bibinfo
			{title} {Understanding Molecular Simulation: From Algorithms to
				Applications}}},\ Computational science\ (\bibinfo  {publisher} {Elsevier
		Science},\ \bibinfo {year} {2001})\BibitemShut {NoStop}%
	\bibitem [{\citenamefont {Mito}(2018)}]{Priv_Comm_Mito}%
	\BibitemOpen
	\bibfield  {author} {\bibinfo {author} {\bibfnamefont {M.}~\bibnamefont
			{Mito}},\ }\href@noop {} {}\bibinfo {howpublished} {{Private Communication}}
	(\bibinfo {year} {2018})\BibitemShut {NoStop}%
	\bibitem [{\citenamefont {A~Osborn}(1945)}]{Osborn1945}%
	\BibitemOpen
	\bibfield  {author} {\bibinfo {author} {\bibfnamefont {J.}~\bibnamefont
			{A~Osborn}},\ }\href {\doibase 10.1103/PhysRev.67.351} {\bibfield  {journal}
		{\bibinfo  {journal} {Physical Review}\ }\textbf {\bibinfo {volume} {67}},\
		\bibinfo {pages} {351} (\bibinfo {year} {1945})}\BibitemShut {NoStop}%
	\bibitem [{\citenamefont {Twengstr\"om}\ \emph {et~al.}(2017)\citenamefont
		{Twengstr\"om}, \citenamefont {Bovo}, \citenamefont {Gingras}, \citenamefont
		{Bramwell},\ and\ \citenamefont {Henelius}}]{Micke17}%
	\BibitemOpen
	\bibfield  {author} {\bibinfo {author} {\bibfnamefont {M.}~\bibnamefont
			{Twengstr\"om}}, \bibinfo {author} {\bibfnamefont {L.}~\bibnamefont {Bovo}},
		\bibinfo {author} {\bibfnamefont {M.~J.~P.}\ \bibnamefont {Gingras}},
		\bibinfo {author} {\bibfnamefont {S.~T.}\ \bibnamefont {Bramwell}}, \ and\
		\bibinfo {author} {\bibfnamefont {P.}~\bibnamefont {Henelius}},\ }\href
	{\doibase 10.1103/PhysRevMaterials.1.044406} {\bibfield  {journal} {\bibinfo
			{journal} {Phys. Rev. Materials}\ }\textbf {\bibinfo {volume} {1}},\ \bibinfo
		{pages} {044406} (\bibinfo {year} {2017})}\BibitemShut {NoStop}%
	\bibitem [{\citenamefont {Stewart}\ \emph {et~al.}(2008)\citenamefont
		{Stewart}, \citenamefont {Deen}, \citenamefont {Andersen}, \citenamefont
		{Schober}, \citenamefont {Barth\'el\'emy}, \citenamefont {Hillier},
		\citenamefont {Murani}, \citenamefont {Hayes},\ and\ \citenamefont
		{Lindenau}}]{D7}%
	\BibitemOpen
	\bibfield  {author} {\bibinfo {author} {\bibfnamefont {J.~R.}\ \bibnamefont
			{Stewart}}, \bibinfo {author} {\bibfnamefont {P.~P.}\ \bibnamefont {Deen}},
		\bibinfo {author} {\bibfnamefont {K.~H.}\ \bibnamefont {Andersen}}, \bibinfo
		{author} {\bibfnamefont {H.}~\bibnamefont {Schober}}, \bibinfo {author}
		{\bibfnamefont {J.-F.}\ \bibnamefont {Barth\'el\'emy}}, \bibinfo {author}
		{\bibfnamefont {J.~M.}\ \bibnamefont {Hillier}}, \bibinfo {author}
		{\bibfnamefont {A.~P.}\ \bibnamefont {Murani}}, \bibinfo {author}
		{\bibfnamefont {T.}~\bibnamefont {Hayes}}, \ and\ \bibinfo {author}
		{\bibfnamefont {B.}~\bibnamefont {Lindenau}},\ }\href@noop {} {\bibfield
		{journal} {\bibinfo  {journal} {Journal of Applied Crystallography}\ }\textbf
		{\bibinfo {volume} {42}},\ \bibinfo {pages} {69 } (\bibinfo {year}
		{2008})}\BibitemShut {NoStop}%
	\bibitem [{\citenamefont {Deen}\ \emph {et~al.}(2018)\citenamefont {Deen},
		\citenamefont {I.M.Bakke}, \citenamefont {Edberg}, \citenamefont
		{Fjellv\r{a}g}, \citenamefont {Leffmann}, \citenamefont {Sandberg},\ and\
		\citenamefont {Wildes}}]{dataThatWeTookOnILL}%
	\BibitemOpen
	\bibfield  {author} {\bibinfo {author} {\bibfnamefont {P.}~\bibnamefont
			{Deen}}, \bibinfo {author} {\bibnamefont {I.M.Bakke}}, \bibinfo {author}
		{\bibfnamefont {R.}~\bibnamefont {Edberg}}, \bibinfo {author} {\bibfnamefont
			{H.}~\bibnamefont {Fjellv\r{a}g}}, \bibinfo {author} {\bibfnamefont
			{K.}~\bibnamefont {Leffmann}}, \bibinfo {author} {\bibfnamefont {L.~{\O}.}\
			\bibnamefont {Sandberg}}, \ and\ \bibinfo {author} {\bibfnamefont
			{A.}~\bibnamefont {Wildes}},\ }\href {\doibase URL
		https://doi.ill.fr/10.5291/ILL-DATA.5-32-856} {\bibfield  {journal} {\bibinfo
			{journal} {ILL}\ } (\bibinfo {year} {2018}),\ URL
		https://doi.ill.fr/10.5291/ILL-DATA.5-32-856}\BibitemShut {NoStop}%
	\bibitem [{\citenamefont {{L. \O . Sandberg}}\ \emph
		{et~al.}(2019)\citenamefont {{L. \O . Sandberg}}, \citenamefont {Haubro},
		\citenamefont {Olsen}, \citenamefont {Guthrie}, \citenamefont {more},
		\citenamefont {Lefmann},\ and\ \citenamefont {Deen}}]{cell}%
	\BibitemOpen
	\bibfield  {author} {\bibinfo {author} {\bibnamefont {{L. \O . Sandberg}}},
		\bibinfo {author} {\bibfnamefont {M.~L.}\ \bibnamefont {Haubro}}, \bibinfo
		{author} {\bibfnamefont {M.~A.}\ \bibnamefont {Olsen}}, \bibinfo {author}
		{\bibfnamefont {M.}~\bibnamefont {Guthrie}}, \bibinfo {author} {\bibnamefont
			{more}}, \bibinfo {author} {\bibfnamefont {K.}~\bibnamefont {Lefmann}}, \
		and\ \bibinfo {author} {\bibfnamefont {P.~P.}\ \bibnamefont {Deen}},\
	}\href@noop {} {\bibfield  {journal} {\bibinfo  {journal} {Review of
				Scientific Instruments}\ }\textbf {\bibinfo {volume} {(in preparation)}}
		(\bibinfo {year} {2019})}\BibitemShut {NoStop}%
	\bibitem [{\citenamefont {Sala}\ \emph {et~al.}(2014)\citenamefont {Sala},
		\citenamefont {Gutmann}, \citenamefont {Prabhakaran}, \citenamefont
		{Pomaranski}, \citenamefont {Mitchelitis}, \citenamefont {Kycia},
		\citenamefont {Porter}, \citenamefont {Castelnovo},\ and\ \citenamefont
		{Goff}}]{Sala14}%
	\BibitemOpen
	\bibfield  {author} {\bibinfo {author} {\bibfnamefont {G.}~\bibnamefont
			{Sala}}, \bibinfo {author} {\bibfnamefont {M.~J.}\ \bibnamefont {Gutmann}},
		\bibinfo {author} {\bibfnamefont {D.}~\bibnamefont {Prabhakaran}}, \bibinfo
		{author} {\bibfnamefont {D.}~\bibnamefont {Pomaranski}}, \bibinfo {author}
		{\bibfnamefont {C.}~\bibnamefont {Mitchelitis}}, \bibinfo {author}
		{\bibfnamefont {J.~B.}\ \bibnamefont {Kycia}}, \bibinfo {author}
		{\bibfnamefont {D.~G.}\ \bibnamefont {Porter}}, \bibinfo {author}
		{\bibfnamefont {C.}~\bibnamefont {Castelnovo}}, \ and\ \bibinfo {author}
		{\bibfnamefont {J.~P.}\ \bibnamefont {Goff}},\ }\href@noop {} {\bibfield
		{journal} {\bibinfo  {journal} {{Nature Materials}}\ }\textbf {\bibinfo
			{volume} {{13}}},\ \bibinfo {pages} {{488}} (\bibinfo {year}
		{{2014}})}\BibitemShut {NoStop}%
	\bibitem [{\citenamefont {Marshall}\ and\ \citenamefont
		{Lovesey}(1971)}]{Marshall71}%
	\BibitemOpen
	\bibfield  {author} {\bibinfo {author} {\bibfnamefont {W.}~\bibnamefont
			{Marshall}}\ and\ \bibinfo {author} {\bibfnamefont {S.~W.}\ \bibnamefont
			{Lovesey}},\ }\href@noop {} {\emph {\bibinfo {title} {Theory of thermal
				neutron scattering: the use of neutrons for the investigation of condensed
				matter}}}\ (\bibinfo  {publisher} {Clarendon Press Oxford},\ \bibinfo {year}
	{1971})\BibitemShut {NoStop}%
	\bibitem [{\citenamefont {Blume}(1963)}]{Blume}%
	\BibitemOpen
	\bibfield  {author} {\bibinfo {author} {\bibfnamefont {M.}~\bibnamefont
			{Blume}},\ }\href@noop {} {\bibfield  {journal} {\bibinfo  {journal} {Phys.
				Rev.}\ }\textbf {\bibinfo {volume} {130}},\ \bibinfo {pages} {1670} (\bibinfo
		{year} {1963})}\BibitemShut {NoStop}%
	\bibitem [{\citenamefont {Fennell}\ \emph {et~al.}(2017)\citenamefont
		{Fennell}, \citenamefont {Mangin-Thro}, \citenamefont {Mutka}, \citenamefont
		{Nilsen},\ and\ \citenamefont {Wildes}}]{D7resolution}%
	\BibitemOpen
	\bibfield  {author} {\bibinfo {author} {\bibfnamefont {T.}~\bibnamefont
			{Fennell}}, \bibinfo {author} {\bibfnamefont {L.}~\bibnamefont
			{Mangin-Thro}}, \bibinfo {author} {\bibfnamefont {H.}~\bibnamefont {Mutka}},
		\bibinfo {author} {\bibfnamefont {G.}~\bibnamefont {Nilsen}}, \ and\ \bibinfo
		{author} {\bibfnamefont {A.}~\bibnamefont {Wildes}},\ }\href {\doibase
		10.1016/j.nima.2017.03.024} {\bibfield  {journal} {\bibinfo  {journal}
			{Nuclear Instruments and Methods in Physics Research Section A: Accelerators,
				Spectrometers, Detectors and Associated Equipment}\ }\textbf {\bibinfo
			{volume} {857}} (\bibinfo {year} {2017}),\
		10.1016/j.nima.2017.03.024}\BibitemShut {NoStop}%
\end{thebibliography}
%\bibliographystyle{ieeetr}


\end{document}